\documentclass[conference, romanappendices]{tex/IEEEtran}
\IEEEoverridecommandlockouts
\usepackage{fancyhdr}
\usepackage[switch,columnwise]{lineno}
\usepackage[linesnumbered,ruled,vlined]{algorithm2e}
\usepackage[noend]{algpseudocode}
\usepackage{amsmath}
\usepackage{stmaryrd}
\usepackage{url}
\usepackage{listings}
\usepackage{mdwlist}
\usepackage{pifont}
\usepackage{graphicx}
\usepackage{wrapfig}
\usepackage{fancyvrb}
\usepackage{paralist}
\usepackage{caption}
\usepackage{comment}
\usepackage{bbm}
\usepackage{dsfont}
\usepackage{mathrsfs}
\usepackage[T1]{fontenc}
\usepackage[dvipsnames]{xcolor}
\usepackage{esvect}
\usepackage{array}
\usepackage{multirow}
\usepackage{rotating}
\usepackage{makecell}
\usepackage{amssymb}
\usepackage{hyperref}
\usepackage{amsfonts}
\usepackage{xparse}
\usepackage[export]{adjustbox}
\usepackage{tabularx}
\usepackage{multicol}
\usepackage{acronym}
\usepackage{subcaption}
\usepackage[]{soul}
\usepackage[]{todonotes}
\usepackage{multicol}
\usepackage{float}
\usepackage[framemethod=tikz]{mdframed}
\usepackage{mathpartir}
\usepackage{amsthm}
\usepackage{mfirstuc}
\usepackage{booktabs}
\usepackage{colortbl}
\usepackage{stackengine}
\usepackage{xspace}
\usepackage{afterpage}
\usepackage[nomessages]{fp}
\usepackage[autolanguage]{numprint}
\usepackage{fmtcount}
\usepackage{microtype}

\definecolor{hgreen}{RGB}{160,255,218}
\definecolor{hred}{RGB}{255,204,204}
\definecolor{babypink}{rgb}{0.96, 0.76, 0.76}

\mdfdefinestyle{graybox}{
	roundcorner=1mm,%
	backgroundcolor=gray!30
}

\newtheoremstyle{bfnote}%
{}{}
{}{}
{\bfseries}{}
{ }{\thmname{#1}\thmnumber{ #2}\thmnote{ (#3)}}
\theoremstyle{bfnote}

\graphicspath{{figs/}}

\tikzset{inlinenotestyle/.append style={align=justify}}
\SetNoFillComment
\DontPrintSemicolon

\SetCommentSty{mycommfont}
\SetFuncSty{texttt}

\newcolumntype{C}[1]{>{\centering\arraybackslash}m{#1}}
\newcolumntype{Y}{>{\raggedleft\arraybackslash}X}
\newcolumntype{Z}{>{\centering\arraybackslash}X}

\newenvironment{conditions*}
{\par\vspace{\abovedisplayskip}\noindent
	\tabularx{\columnwidth}{>{$}l<{$} @{${}={}$} >{\raggedright\arraybackslash}X}}
{\endtabularx\par\vspace{\belowdisplayskip}}

\newcommand{\toolname}{\textsc{Sailfish}\xspace}
\newcommand{\oyente}{{\sc Oyente}\xspace}
\newcommand{\teether}{{\sc teether}\xspace}
\newcommand{\rosette}{{\sc Rosette}\xspace}
\newcommand{\explore}{{\sc Explore}\xspace}
\newcommand{\explorer}{{\sc Explorer}\xspace}
\newcommand{\refine}{{\sc Refine}\xspace}
\newcommand{\refiner}{{\sc Refiner}\xspace}
\newcommand{\securify}{{\sc Securify}\xspace}
\newcommand{\sereum}{{\sc Sereum}\xspace}
\newcommand{\slither}{{\sc Slither}\xspace}
\newcommand{\serif}{{\sc SeRIF}\xspace}
\newcommand{\madmax}{{\sc MadMax}\xspace}
\newcommand{\zeus}{{\sc Zeus}\xspace}

\newcommand{\manticore}{{\sc Manticore}\xspace}
\newcommand{\ethbmc}{{\sc EthBMC}\xspace}
\newcommand{\smartscopy}{{\sc SmartScopy}\xspace}
\newcommand{\smartcheck}{{\sc SmartCheck}\xspace}

\newcommand{\ecf}{{\sc Ecfchecker}\xspace}
\newcommand{\soda}{{\sc Soda}\xspace}
\newcommand{\txspector}{{\sc TxSpector}\xspace}

\newcommand{\vandal}{{\sc Vandal}\xspace}
\newcommand{\sys}{{\sc Sys}\xspace}
\newcommand{\woodpecker}{{\sc Woodpecker}\xspace}
\newcommand{\deadline}{{\sc Deadline}\xspace}
\newcommand{\solidity}{{\sc Solidity}\xspace}
\newcommand{\vyper}{{\sc Vyper}\xspace}

\newcommand{\smart}{smart contract}
\newcommand{\ethereum}{Ethereum}

\newcommand{\etherscan}{{\sc Etherscan}\xspace}
\newcommand{\ether}{{Ether}\xspace}

\newcommand{\reentrancy}{{reentrancy}\xspace}

\newcommand{\tod}{{transaction order dependence}\xspace}
\newcommand{\haz}{{hazardous access}\xspace}
\newcommand{\si}{{state-inconsistency}\xspace}
\newcommand{\vsa}{{value-summary analysis}\xspace}
\newcommand{\sdg}{storage dependency graph\xspace}

\newcommand{\mythril}{{\sc Mythril}\xspace}

\newcommand{\etal}{\textit{et. al.}}

\newcommand{\etc}{\textit{etc.}}
\newcommand{\eg}{\textit{e.g.}}
\newcommand{\viz}{\textit{viz.}}
\newcommand{\ie}{\textit{i.e.}}

\let\num\numprint
\npthousandsep{,}

\newcommand{\Line}[1]{\ensuremath{\sf Line~ #1}}
\newcommand{\Tbl}[1]{\ensuremath{\sf Table~\ref{#1}}}
\newcommand{\Fig}[1]{\ensuremath{\sf Figure~\ref{#1}}}
\newcommand{\Sect}[1]{\ensuremath{\sf Section~\ref{#1}}}
\newcommand{\Appen}[1]{\ensuremath{\sf Appendix~\ref{#1}}}

\NewDocumentCommand{\lstp}{m  m O{}}{\ensuremath{\sf Listings~\ref{#1} ~\text{and} ~\ref{#2}}}
\NewDocumentCommand{\Lines}{m  m O{}}{\ensuremath{\sf Line~ #1 ~\text{and} ~#2}}

\stackMath

\newcounter{sqindex}

\newcommand{\defnote}[3]{
	\newcounter{#1}
	
	\expandafter
	\newcommand\csname cmt#1\endcsname[1]{
		\refstepcounter{#1} 
		\textcolor{#3}{\textbf{#2 [\csname the#1\endcsname]:}} %
		\hl{\textbf{##1.}}} %
	
	\expandafter
	\newcommand\csname cmtdel#1\endcsname[2]{
		\refstepcounter{#1} 
		\textcolor{#3}{\textbf{#2 [\csname the#1\endcsname]:} ##2} %
		\st{\textbf{##1}}} %

	\expandafter
	\newcommand\csname #1\endcsname[1]{%
		\refstepcounter{#1}%
		{%
			\todo[color={#3!30},inline]{%
				\textbf{#2 [\csname the#1\endcsname]:}~##1}%
	}}
}

\definecolor{darkorchid}{rgb}{0.6, 0.2, 0.8}
\defnote{dd}{Dipanjan}{magenta}
\defnote{pb}{Priyanka}{blue}
\defnote{yf}{Yu}{darkorchid}

\newcommand{\contract}{\mathcal{P}}

\newcommand{\valEnv}{\delta}
\newcommand{\pathEnv}{\pi}
\newcommand{\union}{\mu}
\newcommand{\func}{\mathcal{F}}
\newcommand{\stmt}{s}

\definecolor{boxgreen}{HTML}{38761d}
\definecolor{boxred}{HTML}{ff0000}
\definecolor{instrblue}{HTML}{cfe2f3}
\definecolor{instrgreen}{HTML}{b6d7a8}
\definecolor{edgered}{HTML}{ff0000}

\newcommand*\emptycirc[1][0.75ex]{\tikz\draw (0,0) circle (#1);} 
\newcommand*\halfcirc[1][0.75ex]{%
	\begin{tikzpicture}
	\draw[fill] (0,0)-- (90:#1) arc (90:270:#1) -- cycle ;
	\draw (0,0) circle (#1);
	\end{tikzpicture}}
\newcommand*\fullcirc[1][0.75ex]{\tikz\fill (0,0) circle (#1);}

\newcommand*\circledtext[1]{\tikz[baseline=(char.base)]{\node[shape=circle,fill,inner sep=1pt] (char) {\textcolor{white}{\small{#1}}};}}
\newcommand\crule[3][black]{\textcolor{#1}{\rule{#2}{#3}}}

\newcommand{\irule}[2]%
   {\mkern-2mu\displaystyle\frac{#1}{\vphantom{,}#2}\mkern-2mu}
\newcommand{\irulelabel}[3]
{
\mkern-2mu
\begin{array}{ll}
\displaystyle\frac{#1}{\vphantom{,}#2} & \!\!\!\!#3
\end{array}
\mkern-2mu
}

\newtheorem{example}{Example}
\newtheorem{definition}{Definition}

\makeatletter
\def\endthebibliography{%
	\def\@noitemerr{\@latex@warning{Empty `thebibliography' environment}}%
	\endlist
}
\makeatother

\makeatletter
\let\old@ps@IEEEtitlepagestyle\ps@IEEEtitlepagestyle
\def\confheader#1{%
	\def\ps@IEEEtitlepagestyle{%
		\old@ps@IEEEtitlepagestyle%
		\def\@oddhead{\strut\hfill#1\hfill\strut}%
		\def\@evenhead{\strut\hfill#1\hfill\strut}%
	}%
	\ps@headings%
}
\makeatother

\newcommand{\contractsCrawled}{91921}
\newcommand{\contractsExcluded}{2068}
\newcommand{\dataset}{89853}
\newcommand{\smallDataset}{73433}
\newcommand{\mediumDataset}{11730}
\newcommand{\largeDataset}{4690}
\newcommand{\timeBudget}{20}

\newcommand{\securifySafeDAO}{72149}
\newcommand{\securifyUnsafeDAO}{6321}
\newcommand{\securifySafeTOD}{59439}
\newcommand{\securifyUnsafeTOD}{19031}
\newcommand{\securifyTimeout}{10581}
\newcommand{\securifyError}{802}

\newcommand{\mythrilSafeDAO}{25705}
\newcommand{\mythrilUnsafeDAO}{3708}
\newcommand{\mythrilTimeout}{59296}
\newcommand{\mythrilError}{1144}

\newcommand{\vandalSafeDAO}{40607}
\newcommand{\vandalUnsafeDAO}{45971}
\newcommand{\vandalTimeout}{1373}
\newcommand{\vandalError}{1902}

\newcommand{\oyenteSafeDAO}{26924}
\newcommand{\oyenteUnsafeDAO}{269}
\newcommand{\oyenteSafeTOD}{23721}
\newcommand{\oyenteUnsafeTOD}{3472}
\newcommand{\oyenteTimeout}{0}
\newcommand{\oyenteError}{62660}

\newcommand{\clintSafeDAO}{83171}
\newcommand{\clintUnsafeDAO}{2076}
\newcommand{\clintSafeTOD}{77692}
\newcommand{\clintUnsafeTOD}{7555}
\newcommand{\clintTimeout}{1211}
\newcommand{\clintError}{3395}

\newcommand{\manualAnalysisDataset}{750}
\newcommand{\allToolsSuccess}{6581}
\newcommand{\groundTruthDAO}{26}
\newcommand{\groundTruthTOD}{110}

\newcommand{\securifyTPDAO}{9}
\newcommand{\securifyFPDAO}{163}
\newcommand{\securifyFNDAO}{17}
\newcommand{\securifyTPTOD}{102}
\newcommand{\securifyFPTOD}{244}
\newcommand{\securifyFNTOD}{8}

\newcommand{\vandalTPDAO}{26}
\newcommand{\vandalFPDAO}{626}
\newcommand{\vandalFNDAO}{0}
\newcommand{\vandalTPTOD}{--}
\newcommand{\vandalFPTOD}{--}
\newcommand{\vandalFNTOD}{--}

\newcommand{\mythrilTPDAO}{7}
\newcommand{\mythrilFPDAO}{334}
\newcommand{\mythrilFNDAO}{19}
\newcommand{\mythrilTPTOD}{--}
\newcommand{\mythrilFPTOD}{--}
\newcommand{\mythrilFNTOD}{--}

\newcommand{\oyenteTPDAO}{8}
\newcommand{\oyenteFPDAO}{16}
\newcommand{\oyenteFNDAO}{18}
\newcommand{\oyenteTPTOD}{71}
\newcommand{\oyenteFPTOD}{116}
\newcommand{\oyenteFNTOD}{39}

\newcommand{\clintTPDAO}{26}
\newcommand{\clintFPDAO}{11}
\newcommand{\clintFNDAO}{0}
\newcommand{\clintTPTOD}{110}
\newcommand{\clintFPTOD}{59}
\newcommand{\clintFNTOD}{0}

\newcommand{\clintOnlyDAO}{401}
\newcommand{\clintOnlyDAOTriage}{88}
\newcommand{\clintOnlyTOD}{721}
\newcommand{\clintOnlyTODTriage}{107}
\newcommand{\zeroDays}{47}

\newcommand{\securifyAnalysisTimeSmall}{85.51}
\newcommand{\securifyAnalysisTimeMedium}{642.22}
\newcommand{\securifyAnalysisTimeLarge}{823.48}
\newcommand{\securifyAnalysisTimeFull}{196.52}

\newcommand{\vandalAnalysisTimeSmall}{16.35}
\newcommand{\vandalAnalysisTimeMedium}{74.77}
\newcommand{\vandalAnalysisTimeLarge}{177.70}
\newcommand{\vandalAnalysisTimeFull}{30.68}

\newcommand{\mythrilAnalysisTimeSmall}{917.99}
\newcommand{\mythrilAnalysisTimeMedium}{1046.80}
\newcommand{\mythrilAnalysisTimeLarge}{1037.77}
\newcommand{\mythrilAnalysisTimeFull}{941.04}

\newcommand{\oyenteAnalysisTimeSmall}{148.35}
\newcommand{\oyenteAnalysisTimeMedium}{521.16}
\newcommand{\oyenteAnalysisTimeLarge}{675.05}
\newcommand{\oyenteAnalysisTimeFull}{183.45}

\newcommand{\clintAnalysisTimeSmall}{9.80}
\newcommand{\clintAnalysisTimeMedium}{80.78}
\newcommand{\clintAnalysisTimeLarge}{246.89}
\newcommand{\clintAnalysisTimeFull}{30.79}
\newcommand{\clintAverageSATime}{21.74}
\newcommand{\clintAverageVSATime}{0.06}
\newcommand{\clintAverageSETime}{39.22}

\newcommand{\clintStaticOnlyDAO}{3391}
\newcommand{\clintStaticOnlyTOD}{14485}
\newcommand{\clintHavocDAO}{2436}
\newcommand{\clintHavocTOD}{10560}

\newcommand{\clintStaticOnlyWarnings}{16835}
\newcommand{\clintSpeedupDataset}{2000}
\newcommand{\clintPathSummaryTimeout}{430}

\definecolor{verylightgray}{rgb}{.99,.99,.99}
\definecolor{burgundy}{rgb}{0.5, 0.0, 0.13}
\definecolor{brown(web)}{rgb}{0.65, 0.16, 0.16}

\lstdefinelanguage{Solidity}{
	keywords=[1]{anonymous, assembly, assert, balance, break, call, callcode, case, catch, class, constant, continue, constructor, contract, debugger, default, delegatecall, delete, do, else, emit, event, experimental, export, external, false, finally, for, function, gas, if, implements, import, in, indexed, instanceof, interface, internal, is, length, library, log0, log1, log2, log3, log4, memory, modifier, new, payable, pragma, private, protected, public, pure, push, require, return, returns, revert, selfdestruct, send, solidity, storage, struct, suicide, super, switch, then, this, throw, transfer, true, try, typeof, using, value, view, while, with, addmod, ecrecover, keccak256, mulmod, ripemd160, sha256, sha3, , mapping}, %
	keywordstyle=[1]\color{blue},
	keywords=[2]{address, bool, byte, bytes, bytes1, bytes2, bytes3, bytes4, bytes5, bytes6, bytes7, bytes8, bytes9, bytes10, bytes11, bytes12, bytes13, bytes14, bytes15, bytes16, bytes17, bytes18, bytes19, bytes20, bytes21, bytes22, bytes23, bytes24, bytes25, bytes26, bytes27, bytes28, bytes29, bytes30, bytes31, bytes32, enum, int, int8, int16, int24, int32, int40, int48, int56, int64, int72, int80, int88, int96, int104, int112, int120, int128, int136, int144, int152, int160, int168, int176, int184, int192, int200, int208, int216, int224, int232, int240, int248, int256, string, uint, uint8, uint16, uint24, uint32, uint40, uint48, uint56, uint64, uint72, uint80, uint88, uint96, uint104, uint112, uint120, uint128, uint136, uint144, uint152, uint160, uint168, uint176, uint184, uint192, uint200, uint208, uint216, uint224, uint232, uint240, uint248, uint256, var, void, ether, finney, szabo, wei, days, hours, minutes, seconds, weeks, years},	%
	keywordstyle=[2]\color{teal},
	keywords=[3]{block, blockhash, coinbase, difficulty, gaslimit, number, timestamp, msg, data, gas, sig, value, now, tx, gasprice, origin},	%
	keywordstyle=[3]\color{violet} ,
	identifierstyle=\color{black},
	sensitive=false,
	comment=[l]{//},
	morecomment=[s]{/*}{*/},
	commentstyle=\color{brown(web)}\ttfamily,
	stringstyle=\color{red}\ttfamily,
	morestring=[b]',
	morestring=[b]"
}

\lstdefinelanguage{llvm}{
	morecomment = [l]{;},
	morecomment=[l]{//},
	morestring=[b]", 
	sensitive = true,
	classoffset=0,
	commentstyle=\itshape\color{blue!100!black},       %
	morekeywords={
		define, declare, global, constant,
		internal, external, private,
		linkonce, linkonce_odr, weak, weak_odr, appending,
		common, extern_weak,
		thread_local, dllimport, dllexport,
		hidden, protected, default,
		except, deplibs,
		volatile, fastcc, coldcc, cc, ccc,
		x86_stdcallcc, x86_fastcallcc,
		ptx_kernel, ptx_device,
		signext, zeroext, inreg, sret, nounwind, noreturn,
		nocapture, byval, nest, readnone, readonly, noalias, uwtable,
		inlinehint, noinline, alwaysinline, optsize, ssp, sspreq,
		noredzone, noimplicitfloat, naked, alignstack,
		module, asm, align, tail, to,
		addrspace, section, alias, sideeffect, c, gc,
		target, datalayout, triple,
		blockaddress,
		@malloc,@strncpy,@tmpfile,@foo
	},
	classoffset=1, keywordstyle=\color{purple},
	morekeywords={
		fadd, sub, fsub, mul, fmul,
		sdiv, udiv, fdiv, srem, urem, frem,
		and, or, xor,
		icmp, fcmp,
		eq, ne, ugt, uge, ult, ule, sgt, sge, slt, sle,
		oeq, ogt, oge, olt, ole, one, ord, ueq, ugt, uge,
		ult, ule, une, uno,
		nuw, nsw, exact, inbounds,
		phi, call, select, shl, lshr, ashr, va_arg,
		trunc, zext, sext,
		fptrunc, fpext, fptoui, fptosi, uitofp, sitofp,
		ptrtoint, inttoptr, bitcast,
		ret, br, indirectbr, switch, invoke, unwind, unreachable,
		malloc, alloca, free, load, store, getelementptr,
		extractelement, insertelement, shufflevector,
		extractvalue, insertvalue,
	},
	alsoletter={\%},
}

\confheader{%
	\parbox{25cm}{\color{red} To appear in the IEEE Symposium on Security \& Privacy, May 2022. Please cite the conference version of the paper.}
}

\begin{document}
	\title{\toolname: Vetting Smart Contract State-Inconsistency Bugs in Seconds}
	\author{
		\IEEEauthorblockN{
			Priyanka Bose,
			Dipanjan Das,
			Yanju Chen,
			Yu Feng,
			Christopher Kruegel, and
			Giovanni Vigna
		}
	
		\IEEEauthorblockA{University of California, Santa Barbara}
		\IEEEauthorblockA{\{priyanka, dipanjan, yanju, yufeng, chris, vigna\}@cs.ucsb.edu}
	}

	\microtypesetup{disable} %
	\maketitle
	\begin{abstract}
This paper presents \toolname, a scalable system for automatically finding state-inconsistency bugs in smart contracts.
To make the analysis tractable, we introduce a hybrid approach that includes (i) a light-weight exploration phase that dramatically reduces the number of instructions to analyze, and (ii) a precise refinement phase based on symbolic evaluation guided by our novel \vsa, which generates extra constraints to over-approximate the side effects of whole-program execution, thereby ensuring the precision of the symbolic evaluation.
We developed a prototype of \toolname and evaluated its ability to detect two \si flaws, \viz, \reentrancy and \tod (TOD) in \ethereum{} \smart s.
\looseness=-1

Our experiments demonstrate the efficiency of our hybrid approach as well as the benefit of the value summary analysis.
In particular, we show that \toolname outperforms five state-of-the-art \smart{} analyzers (\securify, \mythril, \oyente, \sereum and \vandal) in terms of performance, and precision.
In total, \toolname discovered $\zeroDays$ previously unknown vulnerable smart contracts out of  $\num{\dataset}$ \smart s from \etherscan.
\looseness=-1
\end{abstract}

	\section{Introduction}
\label{introduction}
Smart contracts are programs running on top of the \ethereum{} blockchain.
Due to the convenience of high-level programming languages like \solidity and the security guarantees from the underlying consensus protocol, smart contracts have seen widespread adoption, with over 45 million~\cite{etherscan} instances covering financial products~\cite{case1}, online gaming~\cite{case2}, real estate, and logistics.
Consequently, a vulnerability in a contract can lead to tremendous losses, as demonstrated by recent attacks~\cite{attack1,attack2,attack3,attack4}.
For instance, the notorious ``TheDAO''~\cite{dao-attack} \reentrancy attack led to a financial loss of about \$50M in 2016.
Furthermore, in recent years, several other \reentrancy attacks, \eg,  Uniswap~\cite{uniswap-attack},  Burgerswap~\cite{burgerswap-attack}, Lendf.me~\cite{lendf-attack}, resulted in multimillion dollar losses.
To make things worse, smart contracts are \emph{immutable}---once deployed, the design of the consensus protocol makes it particularly difficult to fix bugs.
Since \smart s are not easily upgradable, auditing the contract's source pre-deployment, and deploying a bug-free contract is even more important than in the case of traditional software.
\looseness=-1

In this paper, we present a scalable technique to detect \emph{\si} (SI) bugs---a class of vulnerabilities that enables an attacker to manipulate the global state, \ie, the storage variables of a contract, by tampering with either the order of execution of multiple transactions (\emph{\tod} (TOD)), or the control-flow inside a single transaction (\emph{\reentrancy}).
In those attacks, an attacker can tamper with the critical storage variables that transitively have an influence on money transactions through data or control dependency.
Though ``TheDAO''~\cite{dao-attack} is the most well-known attack of this kind, through an offline analysis~\cite{txspector,sereum} of the historical on-chain data, researchers have uncovered several instances of past attacks that leveraged \si vulnerabilities. 
\looseness=-1

While there are existing tools for detecting vulnerabilities due to state-inconsistency bugs, they either aggressively over-approximate the execution of a smart contract, and report false alarms~\cite{securify,madmax}, or they precisely enumerate~\cite{mythril,oyente} concrete or symbolic traces of the entire smart contract, and hence, cannot scale to large contracts with many paths.
Dynamic tools~\cite{sereum,txspector} scale well, but can detect a \si bug only when the evidence of an active attack is present.
Moreover, existing tools adopt a syntax-directed pattern matching that may miss bugs due to incomplete support for potential attack patterns~\cite{securify}.

A static analyzer for \si bugs is crucial for pre-deployment auditing of smart contracts, but designing such a tool comes with its unique set of challenges.
For example, a \smart{} exposes public methods as interfaces to interact with the outside world.
Each of these methods is an entry point to the contract code, and can potentially alter the persistent state of the contract by writing to the storage variables.
An attacker can invoke \textit{any} method(s), \textit{any} number of times, in \textit{any} arbitrary order---each invocation potentially impacting the overall contract state.
Since different contracts can communicate with each other through public methods, it is even harder to detect a cross-function attack where the attacker can stitch calls to multiple public methods to launch an attack.
Though \sereum~\cite{sereum} and \ecf~\cite{ecf} detect cross-function attacks, they are dynamic tools that reason about one single execution.
However, statically detecting \si bugs boils down to reasoning about the entire contract control and data flows, over multiple executions.
This presents significant scalability challenges, as mentioned in prior work~\cite{sereum}.\looseness=-1

This paper presents \toolname, a highly scalable tool that is aimed at automatically identifying \si bugs in \smart s.
To tackle the scalability issue associated with statically analyzing a contract, \toolname adopts a hybrid approach that combines a light-weight \explore phase, followed by a \refine phase guided by our novel \emph{\vsa}, which constrains the scope of storage variables.
Our \explore phase dramatically reduces the number of relevant instructions to reason about, while the \vsa in the \refine phase further improves performance while maintaining the precision of symbolic evaluation.
Given a smart contract, \toolname first introduces an \explore phase that converts the contract into a \emph{\sdg} (SDG) $G$.
This graph summarizes the side effects of the execution of a contract on storage variables in terms of read-write dependencies.
State-inconsistency vulnerabilities are modeled as graph queries over the SDG structure.
A vulnerability query returns either an empty result---meaning that the contract is not vulnerable, or a potentially vulnerable subgraph $g$ inside $G$ that matches the query.
In the second case, there are two possibilities: either the contract is indeed vulnerable, or $g$ is a false alarm due to the over-approximation of the static analysis. 

To prune potential false alarms, \toolname leverages a \refine phase based on symbolic evaluation.
However, a conservative symbolic executor would initialize the storage variables as \textit{unconstrained}, which would, in turn, hurt the tool's ability to prune many infeasible paths.
To address this issue, \toolname incorporates a light-weight \emph{\vsa} (VSA) that summarizes the value constraints of the storage variables, which are used as the pre-conditions of the symbolic evaluation.
Unlike prior summary-based approaches~\cite{solar,Godefroid07,AnandGT08} that compute summaries path-by-path, which results in full summaries (that encode all bounded paths through a procedure), leading to scalability problems due to the exponential growth with procedure size, 
our VSA summarizes \emph{all paths} through a finite (loop-free) procedure, and it produces compact (polynomially-sized) summaries.
As our evaluation shows, VSA not only enables \toolname to refute more false positives, but also scales much better to large contracts compared to a classic summary-based symbolic evaluation strategy.
\looseness=-1

We evaluated \toolname on the entire data set from \etherscan~\cite{etherscan} ($\num{\dataset}$ contracts), and showed that our tool is efficient and effective in detecting \si bugs.
\toolname significantly outperforms all five state-of-the-art \smart{} analyzers we evaluated against, in the number of reported false positives and false negatives.
For example, on average \toolname took only $\clintAnalysisTimeFull$ seconds to analyze a \smart, which is $\FPeval{\v}{round(\mythrilAnalysisTimeFull/\clintAnalysisTimeFull,0)}\v$ times faster than \mythril~\cite{mythril}, and $\FPeval{\v}{round(\securifyAnalysisTimeFull/\clintAnalysisTimeFull,0)}\text{\numberstringnum{\v}}$ orders of magnitude faster than \securify~\cite{securify}. 

In summary, this paper makes the following contributions: 

\begin{itemize}
\item We define \si{} vulnerabilities and identify two of its root-causes (\Sect{motivation}),
including a new \reentrancy attack pattern that has not been investigated in the previous literature.

\item We model \si{} detection as \haz queries over a unified, compact graph representation (called a \emph{\sdg} (SDG)), which encodes the high-level semantics of smart contracts over global states. (\Sect{static_analysis})

\item We propose a novel \emph{\vsa} that efficiently computes global constraints over storage variables, which 
when combined with symbolic evaluation, enables \toolname to significantly reduce false alarms. (\Sect{refinement})

\item We perform a systematic evaluation of \toolname on the entire data set
from \etherscan.
Not only does \toolname outperforms state-of-the-art \smart{} analyzers in terms of both run-time and precision, but also is able to uncover $\zeroDays$ zero-day vulnerabilities (out of $\FPeval{\v}{round(\clintOnlyDAOTriage+\clintOnlyTODTriage,0)}\v$ contracts that we could manually analyze) not detected by any other tool. (\Sect{evaluation})

\item In the spirit of open science, we pledge to release both the tool and the experimental data to further future research.
\end{itemize}

	\section{Background} 
\label{background}

This section introduces the notion of the state of a \smart, and provides a brief overview of the vulnerabilities leading to an inconsistent state during a contract's execution.

\noindent
\textbf{Smart contract.}
\ethereum{} \smart s are written in high-level languages like \solidity, \vyper, \etc, and are compiled down to the EVM (\ethereum{} Virtual Machine) bytecode.
Public/external methods of a contract, which act as independent entry points of interaction, can be invoked in two ways: either by a \textit{transaction}, or from another contract.
We refer to the invocation of a public/external method from outside the contract as an \textit{event}.
Note that events exclude method calls originated from inside the contract, \ie, a method $f$ calling another method $g$.
A \textit{schedule} $\mathcal{H}$ is a valid sequence of events that can be executed by the EVM.
The events of a schedule can originate from one or more transactions.
Persistent data of a contract is stored in the storage variables which are, in turn, recorded in the blockchain.
The \textit{contract state} $\Delta = (\mathcal{V}, \mathcal{B})$ is a tuple, where $\mathcal{V} = \{V_1, V_2, V_3,..., V_n\}$ is the set of all the storage variables of a contract, and $\mathcal{B}$ is its balance.

\noindent
\textbf{State inconsistency (SI).}
When the events of a schedule $\mathcal{H}$ execute on an initial state $\Delta$ of a contract, it reaches the final state $\Delta^\prime$.
However, due to the presence of several sources of non-determinism~\cite{NPChecker} during the execution of a \smart{} on the \ethereum{} network, $\Delta^\prime$ is not always predictable.
For example,
two transactions are not guaranteed to be processed in the order in which they got scheduled.
Also, an external call $e$ originated from a method $f$ of a contract $\mathcal{C}$ can transfer control to a malicious actor, who can now subvert the original control and data-flow by re-entering $\mathcal{C}$ through any public method $f^\prime \in \mathcal{C}$ in the same transaction, even before the execution of $f$ completes.
Let $\mathcal{H}_1$ be a schedule that does not exhibit any of the above-mentioned non-deterministic behavior.
However, due to either reordering of transactions, or reentrant calls, it might be possible to rearrange the events of $\mathcal{H}_1$ to form another schedule $\mathcal{H}_2$.
If those two schedules individually operate on the same initial state $\Delta$, but yield different final states, we consider the contract to have a \si.

\noindent
\textbf{Reentrancy.}
If a contract $\mathcal{A}$ calls another contract $\mathcal{B}$, the \ethereum{} protocol allows $\mathcal{B}$ to call back to any public/external method $m$ of $\mathcal{A}$ in the same transaction before even finishing the original invocation.
An attack happens when $\mathcal{B}$ reenters $\mathcal{A}$ in an inconsistent state before $\mathcal{A}$ gets the chance to update its internal state in the original call.
Launching an attack executes operations that consume gas.
Though, \solidity tries to prevent such attacks by limiting the gas stipend to $\num{2300}$ when the call is made through \texttt{send} and \texttt{transfer} APIs, the \texttt{call} opcode puts no such restriction---thereby making the attack possible.
\looseness=-1

In \Fig{fig:background_reentrancy},
the \texttt{withdraw} method transfers Ethers to a user if their account balance permits, and then updates the account accordingly.
From the external call at \Line{4}, a malicious user (attacker) can reenter %
the \texttt{withdraw} method of the \texttt{Bank} contract.
It makes \Line{3} read a stale value of the account balance, which was supposed to be updated at \Line{5} in the original call.
Repeated calls to the \texttt{Bank} contract can drain it out of Ethers, because the sanity check on the account balance at \Line{3} never fails.
One such infamous attack, dubbed ``TheDAO'' \cite{dao-attack}, siphoned out over USD \$50 million worth of \ether from a crowd-sourced contract in 2016.

Though the example presented above depicts a typical \reentrancy attack scenario, such attacks can occur in a more convoluted setting, \eg, \textit{cross-function}, \textit{create-based}, and \textit{delegate-based}, as studied in prior work~\cite{sereum}.
A \textit{cross-function} attack spans across multiple functions.
For example, a function $f_1$ in the victim contract $\mathcal{A}$ issues an untrusted external call, which transfers the control over to the attacker  $\mathcal{B}$.
In turn, $\mathcal{B}$ reenters $\mathcal{A}$, but through a different function $f_2$.
A \textit{delegate-based} attack happens when the victim contract $\mathcal{A}$ delegates the control to another contract $\mathcal{C}$, where contract $\mathcal{C}$ issues an untrusted external call.
In case of a \textit{create-based} attack, the victim contract $\mathcal{A}$ creates a new child contract $\mathcal{C}$, which issues an untrusted external call inside its constructor.

\noindent
\textbf{Transaction Order Dependence (TOD).}
Every \ethereum{} transaction specifies the upper limit of the \textit{gas} amount one is willing to spend on that transaction.
Miners choose the ones offering the most incentive for their mining work, thereby inevitably making the transactions offering lower \textit{gas} starve for an indefinite amount of time.
By the time a transaction $T_1$ (scheduled at time $t_1$) is picked up by a miner, the network and the contract states might change due to another transaction $T_2$ (scheduled at time $t_2$) getting executed beforehand, though $t_1 < t_2$.
This is known as Transaction Order Dependence (TOD) \cite{tod-attack}, or \textit{front-running} attack.
\Fig{fig:background_tod}  features a queuing system where an user can reserve a slot (\Line{3, 4}) by submitting a transaction.
An attacker can succeed in getting that slot by eavesdropping on the gas limit set by the victim transaction, and incentivizing the miner by submitting a transaction with a higher gas limit.
Refer to \Sect{si_bugs} where we connect \reentrancy and TOD bugs to our notion of \si.
\vspace{-2mm}
\begin{figure}[t]
	\centering
	\begin{subfigure}[t]{0.49\columnwidth}
\begin{lstlisting}[
			caption=,
			stepnumber=1,
			firstnumber=1,
			basicstyle=\tiny,
%			linebackgroundcolor=
%			{\ifnum\value{lstnumber}=5\color{babypink}
%				\else\fi},
			language=Solidity]
contract Bank {    
 function withdraw(uint amount){
   if(accounts[msg.sender] >= amount){
        msg.sender.call.value(amount);
        accounts[msg.sender] -= amount;
   }
 }
}
\end{lstlisting}
			\vspace{-6.5mm}
			\caption{}
			\label{fig:background_reentrancy}
	\end{subfigure}\hfill
	~ 
	\begin{subfigure}[t]{0.45\columnwidth}
		\centering
\begin{lstlisting}[
		caption=,
		stepnumber=1,
		firstnumber=1,
		basicstyle=\tiny,
%		linebackgroundcolor=
%		{\ifnum\value{lstnumber}=3\color{babypink}
%			\else\fi},
		language=Solidity]
contract Queue {
  function reserve(uint256 slot){
	if (slots[slot] == 0) {
		slots[slot] = msg.sender;
	}
  }
}
\end{lstlisting}
		\vspace{-2mm}
		\caption{}
		\label{fig:background_tod}
		
	\end{subfigure}
	\vspace{-2.5mm}
	\caption{\small In \Fig{fig:background_reentrancy}, the \texttt{accounts} mapping is updated after the external call at \Line{4} .
	This allows the malicious caller to reenter the \texttt{withdraw()} function in an inconsistent state. 
	\Fig{fig:background_tod} presents a contract that implements a queuing system that reserves slots on a  first-come-first-serve basis leading to a potential TOD attack.}
	\vspace{-6mm}
\end{figure}

	\vspace{2mm}
\section{Motivation} 
\label{motivation}
This section introduces motivating examples of \si (SI) vulnerabilities, 
the challenges associated with automatically detecting them, how state-of-the-art techniques fail to tackle those challenges, and our solution.

\subsection{\textbf{Identifying the root causes of SI vulnerabilities}}
By manually analyzing prior instances of \reentrancy and TOD bugs---two popular SI vulnerabilities (\Sect{background}), and the warnings emitted by the existing automated analysis tools~\cite{sereum, mythril, securify, oyente}, we observe that an SI vulnerability occurs when the following preconditions are met:
\textbf{(i)} two method executions, or transactions---both referred to as \textit{threads} ($th$)---operate on the same storage state, and 
\textbf{(ii)} either of the two happens---\textbf{(a)} \textbf{Stale Read (SR)}: 
The attacker thread $th_a$ diverts the flow of execution to read a stale value from \texttt{storage($v$)} before the victim thread $th_v$ gets the chance to legitimately update the same in its flow of execution.
The \reentrancy vulnerability presented in \Fig{fig:background_reentrancy} is the result of a stale read.
\textbf{(b)} \textbf{Destructive Write (DW)}: 
The attacker thread $th_a$ diverts the flow of execution to preemptively write to \texttt{storage($v$)} before the victim thread $th_v$ gets the chance to legitimately read the same in its flow of execution.
The TOD vulnerability presented in \Fig{fig:background_tod} is the result of a destructive write.\looseness=-1

While the SR pattern is well-studied in the existing literature~\cite{securify,sereum,oyente,vandal}, and detected by the respective tools with varying degree of accuracy, the \reentrancy attack induced by the DW pattern 
has never been explored by the academic research community.
Due to its conservative strategy of flagging any state access following an external call without considering if it creates an inconsistent state, \mythril raises alarms for a super-set of DW patterns, leading to a high number of false positives.
In this work, we not only identify the root causes of SI vulnerabilities, but also unify the detection of both the patterns with the notion of \haz (\Sect{motivation}).\looseness=-1

\subsection{\textbf{Running examples}}
\noindent
\textbf{Example 1.}
The contract in \Fig{fig:false_negative} is vulnerable to \reentrancy due to destructive write.
It allows for the splitting of funds held in the payer's account between two payees --- \texttt{a} and \texttt{b}.
For a payer with id \texttt{id}, \texttt{updateSplit} records the fraction ($\%$) of her fund to be sent to the first payer in \texttt{splits[id]} (\Line{5}) .
In turn, \texttt{splitFunds} transfers \texttt{splits[id]} fraction of the payer's total fund to payee \texttt{a}, and the remaining to payee \texttt{b}.
Assuming that the payer with $\texttt{id}=0$ is the attacker, she executes the following sequence of calls in a transaction --
\textbf{(1)} calls $\texttt{updateSplit(0,100)}$ to set payee \texttt{a}'s split to $100\%$ (\Line{5});
\textbf{(2)} calls $\texttt{splitFunds(0)}$ to transfer her entire fund to payee \texttt{a} (\Line{16});
\textbf{(3)} from the fallback function, reenters $\texttt{updateSplit(0,0)}$ to set payee \texttt{a}'s split to $0\%$ (\Line{5});
\textbf{(4)} returns to \texttt{splitFunds} where her entire fund is \textit{again} transferred (\Line{19}) to payee \texttt{b}.
Consequently, the attacker is able to trick the contract into double-spending the amount of Ethers held in the payer's account.\looseness=-1
\begin{figure}[t]
\begin{lstlisting}[
		caption=,
		stepnumber=1,
		firstnumber=1,
		basicstyle=\ttfamily\scriptsize,
%		linebackgroundcolor=
%		{\ifnum\value{lstnumber}=5\color{babypink}
%		\else\ifnum\value{lstnumber}=16\color{babypink}
%		\else\ifnum\value{lstnumber}=19\color{babypink}	
%		\else\fi\fi\fi},
		language=Solidity]
// [Step 1]: Set split of 'a' (id = 0) to 100(%)
// [Step 4]: Set split of 'a' (id = 0) to 0(%)
function updateSplit(uint id, uint split) public{
   require(split <= 100);
   splits[id] = split;
}

function splitFunds(uint id) public {
   address payable a = payee1[id];
   address payable b = payee2[id];
   uint depo = deposits[id];
   deposits[id] = 0;

   // [Step 2]: Transfer 100% fund to 'a'
   // [Step 3]: Reenter updateSplit
   a.call.value(depo * splits[id] / 100)(""); 

   // [Step 5]: Transfer 100% fund to 'b'
   b.transfer(depo * (100 - splits[id]) / 100);
}
\end{lstlisting}
	\vspace{-0.15in}
	\caption{\small The attacker reenters \texttt{updateSplit} from the external call at \Lines{16} and sets \texttt{splits[id] = 0}.
		This enables the attacker to transfer all the funds again to \texttt{b}.}
	\label{fig:false_negative}
	\vspace{-7mm}
\end{figure}

\noindent
\textbf{Example 2.}
The contract in \Fig{fig:false_positive} is non-vulnerable (safe).
The \texttt{withdrawBalance} method allows the caller to withdraw funds from her account.
The storage variable \texttt{userBalance} is updated (\Line{10}) after the external call (\Line{9}).
In absence of the \texttt{mutex}, the contract could contain a \reentrancy bug due to the delayed update. 
However, the \texttt{mutex} is set to \texttt{true} when the function is entered the first time.
If an attacker attempts to reenter \texttt{withdrawBalance} from her fallback function, the check at \Line{4} will foil such an attempt.
Also, the \texttt{transfer} method adjusts the account balances of a sender and a receiver, and is not reentrant due to the same reason (\texttt{mutex}).\looseness=-1

\subsection{\textbf{State of the vulnerability analyses}}
In light of the examples above, we outline the key challenges 
encountered by the state-of-the-art techniques, \ie, \securify~\cite{securify}, \vandal~\cite{vandal}, \mythril~\cite{mythril}, \oyente~\cite{oyente}, and \sereum~\cite{sereum} that find \si (SI) vulnerabilities.
\Tbl{tbl:tool_comparison} summarizes our observations.

\noindent
\textbf{Cross-function attack.}
The public methods in a \smart{} act as independent entry points.
Instead of reentering the same function, as in the case of a traditional \reentrancy attack,
in a cross-function attack, the attacker can reenter the contract through any public function.
Detecting cross-function vulnerabilities poses a significantly harder challenge than single-function \reentrancy, because every external call can jump back to any public method---leading to an explosion in the search space due to a large number of potential call targets.

Unfortunately, most of the state of the art techniques cannot detect cross-function attacks.
For example, the \textit{No Write After Call} (\texttt{NW}) strategy of \securify{} identifies a storage variable write (\texttt{SSTORE}) following a \texttt{CALL} operation as a potential violation.
\mythril{} adopts a similar policy, except it also warns when a state variable is read after an external call.
Both \vandal{} and \oyente{} check if a \texttt{CALL} instruction at a program point can be reached by a recursive call to the enclosing function.
In all four tools, \reentrancy is modeled after The DAO~\cite{dao-attack} attack, and therefore scoped within a single function.
Since the attack demonstrated in Example~1 spans across both the \texttt{updateSplit} and \texttt{splitFunds} methods, detecting such an attack is out of scope for these tools.
Coincidentally, the last three tools raise alarms here for the wrong reason, due to the over-approximation in their detection strategies.
\sereum is a run-time bug detector that detects cross-function attacks.
When a transaction returns from an external call, \sereum write-locks all the storage variables that influenced control-flow decisions in any previous invocation of the contract during the external call.
If a locked variable is re-written going forward, an attack is detected.
\sereum fails to detect the attack in Example~1 (\Fig{fig:false_negative}), because it would not set any lock due to the absence of any control-flow deciding state variable~\footnote{A recent extension~\cite{sereum-repo} of \sereum adds support for unconditional \reentrancy attacks by tracking data-flow dependencies.
However, they only track data-flows from storage variables to the parameters of calls.
As a result, even with this extension, \sereum would fail to detect the attack in Example~1.}.\looseness=-1

\textit{Our solution}: To mitigate the state-explosion issue inherent in static techniques, \toolname{} performs a taint analysis from the arguments of a public method to the \texttt{CALL} instructions to consider only those external calls where the destination can be controlled by an attacker.
Also, we keep our analysis tractable by analyzing public functions in \textit{pairs}, instead of modeling an arbitrarily long call-chain required to synthesize exploits.

\begin{figure}[t]
		{ \centering
\begin{lstlisting}[
			caption=,
			basicstyle=\ttfamily\scriptsize,
			stepnumber=1,
			firstnumber=1,
%			linebackgroundcolor=
%			{\ifnum\value{lstnumber}=4\color{babypink}
%			\else\ifnum\value{lstnumber}=17\color{babypink}	
%			\else\fi\fi},
			language=Solidity]
function withdrawBalance(uint amount) public {
    //[Step 1]: Enter when mutex is false
    //[Step 4]: Early return, since mutex is true
    if (mutex == false) {
      //[Step 2]: mutex = true prevents re-entry
      mutex = true;
      if (userBalance[msg.sender] > amount) {
         //[Step 3]: Attempt to reenter
         msg.sender.call.value(amount)("");
         userBalance[msg.sender] -= amount;
      }
      mutex = false;
    }
}

function transfer(address to, uint amt) public {
    if (mutex == false) {
        mutex = true;
        if (userBalance[msg.sender] > amt) {
            userBalance[to] += amt;
            userBalance[msg.sender] -= amt;
        }
        mutex = false;
    }
}   
\end{lstlisting}
		}
	\vspace{-0.15in}
	\caption{\small \Line{6} sets \texttt{mutex} to \texttt{true}, which prohibits an attacker from reentering
	by invalidating the path condition (\Line{4}).}
	\label{fig:false_positive}
	\vspace{-6mm}
\end{figure}

\noindent
\textbf{Hazardous access.}
Most tools apply a conservative policy, and report a read/write from/to a state variable following an external call as a possible \reentrancy attack.
Since this pattern alone is not sufficient to lead the contract to an inconsistent state, they generate a large number of false positives.
Example~1 (\Fig{fig:false_negative}) without the \texttt{updateSplit} method is \textit{not} vulnerable, since \texttt{splits[id]} cannot be modified any more.
However, \mythril, \oyente{}, and \vandal{} flag the modified example as vulnerable, due to the conservative detection strategies they adopt, as discussed before.

\textit{Our solution}:
We distinguish between \textit{benign} and \textit{vulnerable} reentrancies, \ie, \reentrancy as a feature \textit{vs.} a bug.
We only consider \reentrancy to be vulnerable if it can be leveraged to induce a \si (SI).
Precisely, if two operations 
\textbf{(a)} operate on the same state variable,
\textbf{(b)} are reachable from public methods, and
\textbf{(c)} at-least one is a \texttt{write}---we call these two operations a \haz pair. 
The notion of \haz unifies both Stale Read (SR), and Destructive Write (DW).
\toolname performs a lightweight static analysis to detect such hazardous accesses.
Since the modified Example~1 (without the \texttt{updateSplit}) presented above does not contain any hazardous access pair, we do not flag it as vulnerable.\looseness=-1

\begin{table}
   	\centering
   	\scriptsize
   	\caption{Comparison of smart-contract bug-finding tools.}
   	\begin{tabular}
   		{C{0.30\columnwidth}	%
   			C{0.10\columnwidth}		%
   			C{0.10\columnwidth}		%
   			C{0.10\columnwidth}		%
   			C{0.10\columnwidth}		%
   			C{0.10\columnwidth}}	%
   		
   		\toprule
   		\textbf{Tool} & \textbf{Cr.} & \textbf{Haz.} & \textbf{Scl.} & \textbf{Off.} \\
   		\midrule
   		
   		\rowcolor{black!10} \securify~\cite{securify} & \emptycirc & \emptycirc & \halfcirc & \fullcirc \\
   		
   		\vandal~\cite{vandal} & \emptycirc & \emptycirc & \fullcirc & \fullcirc \\
   		
   		\rowcolor{black!10} \mythril~\cite{mythril} & \emptycirc & \emptycirc & \emptycirc & \fullcirc \\
   		
   		\oyente~\cite{oyente} & \emptycirc & \emptycirc & \halfcirc & \fullcirc \\
   		
   		\rowcolor{black!10} \sereum~\cite{sereum} & \fullcirc & \emptycirc & \fullcirc & \emptycirc \\
   		
   		\toolname & \fullcirc & \fullcirc & \fullcirc & \fullcirc \\
   		
   		\bottomrule
   	\end{tabular}
   	\caption*{\small \fullcirc[0.75ex] Full \halfcirc[0.75ex] Partial \emptycirc[0.75ex] No support. \textbf{Cr.}: Cross-function, \textbf{Haz.}: Hazardous access, \textbf{Scl.}: Scalability, \textbf{Off.}: Offline detection}
   	\label{tbl:tool_comparison}
   	\vspace{-8mm}
\end{table}

\noindent

\noindent
\textbf{Scalability.}
Any \solidity{} method marked as either \texttt{public} or \texttt{external} can be called by an external entity \textit{any} number of times in \textit{any} arbitrary order---which
translates to an unbounded search space during static reasoning.
\securify~\cite{securify} relies on a \texttt{Datalog}-based data-flow analysis, which might fail to reach a fixed point in a reasonable amount of time, as the size of the contract grows.
\mythril~\cite{mythril} and \oyente~\cite{oyente} are symbolic-execution-based tools that share the common problems suffered by any symbolic engine.

\textit{Our solution}:
In \toolname, the symbolic verifier validates a program path involving hazardous accesses.
Unfortunately, the path could access state variables that are likely to be used elsewhere in the contract.
It would be very expensive for a symbolic checker to perform a whole-contract analysis required to precisely model
those state variables.
We augment the verifier with a \textit{value summary} that over-approximates the side-effects of the public methods on the state variables across all executions.
This results in an inexpensive symbolic evaluation that conservatively prunes false positives.

\noindent
\textbf{Offline bug detection.}
Once deployed, a contract becomes immutable.
Therefore, it is important to be able to detect bugs prior to the deployment.
However, offline (static) approaches come with their unique challenges.
Unlike an online (dynamic) tool that detects an ongoing attack in just one execution, a static tool needs to reason about all possible combinations of the contract's public methods while analyzing SI issues.
As a static approach, \toolname{} needs to tackle all these challenges.

\subsection{\textbf{\toolname overview}}
\begin{figure}[t]
	\vspace{-2mm}
	\centering
	\includegraphics[width=\columnwidth]{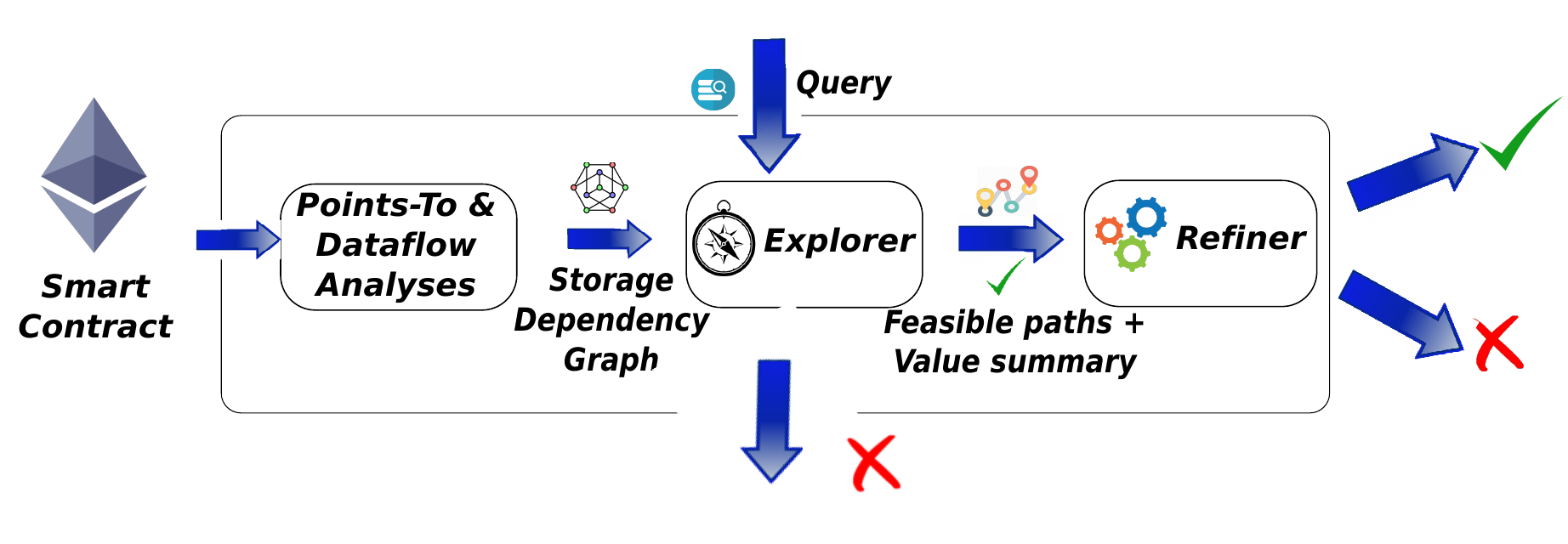}
	\vspace{-10mm}
	\caption{Overview of \toolname}
	\label{fig:overview}
	\vspace{-6.5mm}
\end{figure}

This section provides an overview (\Fig{fig:overview}) of \toolname which consists of the \explorer and the \refiner modules.

\noindent
\textbf{Explorer.}
From a contract's source, \toolname{} statically builds a \textit{\sdg} (SDG) (\Sect{sec:sdg}) which over-approximates the read-write accesses (\Sect{sec:hazardous_access}) on the storage variables along all possible execution paths.
State-inconsistency (SI) vulnerabilities are modeled as graph queries over the SDG.
If the query results in an empty set, the contract is certainly non-vulnerable.
Otherwise, we generate a counter-example which is subject to further validation by the \refiner.

\vspace{-1.5mm}
\begin{example}
Example~1 (\Fig{fig:false_negative}) contains a \reentrancy bug that spans across two functions.
The attacker is able to create an SI by leveraging hazardous accesses---\texttt{splits[id]} influences (read) the argument of the external call at \Line{16} in \texttt{splitFunds}, and it is set (write) at \Line{5} in \texttt{updateSplit}.
The counter-example returned by the \explorer is \circledtext{11} $\rightarrow$ \circledtext{12} $\rightarrow$ \circledtext{16} $\rightarrow$ \circledtext{4} $\rightarrow$ \circledtext{5}.
Similarly, in Example~2 (\Fig{fig:false_positive}), when \texttt{withdrawBalance} is composed with \texttt{transfer} to model a cross-function attack, \toolname{} detects the \textit{write} at \Line{10}, and the \textit{read} at \Line{19} as hazardous.
Corresponding counter-example is \circledtext{4} ... \circledtext{9} $\rightarrow$ \circledtext{17} ... \circledtext{19}.
In both the cases, the \explorer detects a potential SI, so conservatively they are flagged as \textit{possibly vulnerable}.
However, this is incorrect for Example~2.
Thus, we require an additional step to refine the initial results.\looseness=-1
\end{example}

\vspace{-2mm}
\noindent
\textbf{Refiner.}
Although the counter-examples obtained from the \explorer span across only two public functions $\texttt{P}_1$ and $\texttt{P}_2$, the path conditions in the counter-examples may involve state variables that can be operated on by the public methods $\texttt{P}^*$ other than those two.
For example, in case of \reentrancy, the attacker can alter the contract state by invoking $\texttt{P}^*$ after the external call-site---which makes reentry to $\texttt{P}_2$ possible.
To alleviate this issue, we perform a contract-wide \vsa that computes the necessary pre-conditions to set the values of storage variables.
The symbolic verifier consults the value summary when evaluating the path constraints.

\vspace{-2mm}
\begin{example}
In Example~2 (\Fig{fig:false_positive}), the \refiner{} would conservatively assume the \texttt{mutex} to be \textit{unconstrained} after the external call at \Line{9} in absence of a value summary -- which would make the path condition feasible.
However, the summary (\Sect{refinement}) informs the symbolic checker that 
all the possible program flows require the \texttt{mutex} already to be \texttt{false}, in order to set the \texttt{mutex} to \texttt{false} again.
Since the pre-condition conflicts with the program-state $\valEnv = \{\mathtt{mutex}\mapsto \mathtt{true}\}$ (set by \Line{6}), \toolname{} refutes the possibility of the presence of a \reentrancy, thereby pruning the false warning.
\end{example}

	\section{State Inconsistency Bugs}
\label{si_bugs}
In this section, we introduce the notion of \si, and how it is related to \reentrancy and TOD bugs.

Let $\vec{\mathcal{F}}$ be the list of all public/external functions in a contract $\mathcal{C}$ defined later in \Fig{fig:syntax-ir}.
For each function $\mathcal{F} \in \vec{\mathcal{F}}$, we denote $\mathcal{F}.\mathsf{statements}$ to be the statements of $\mathcal{F}$, and $f = \mathcal{F}.\mathsf{name}$ to be the name of $\mathcal{F}$.
In \ethereum, one or more functions can be invoked in a transaction $T$.
Since the contract code is executed by the EVM, the value of its \textit{program counter} (PC) deterministically identifies every statement $s \in \mathcal{F}.\mathsf{statements}$ during run-time.
An \textit{event} $e = \langle pc, f(\vec{x}), inv \rangle$ is a $3$-tuple that represents the $inv$-th invocation of the function $\mathcal{F}$ called from outside (\ie, external to the contract $\mathcal{C}$) with arguments $\vec{x}$.
Identical invocation of a function $\mathcal{F}$ is associated with the same arguments.
For events, we disregard internal subroutine calls, \eg, if the function $\mathcal{F}$ calls another public function $\mathcal{G}$ from inside its body, the latter invocation does not generate an event.
In other words, the notion of events captures the occurrences when a public/external method of a contract is called externally, \ie, across the contract boundary.
Functions in events can be called in two ways: either directly by $T$, or by another contract.
If an external call statement $s_c \in \mathcal{F}_c.\mathsf{statements}$ results in a reentrant invocation of $\mathcal{F}$, then $pc$ holds the value of the program counter of $s_c$.
In this case, we say that the execution of $\mathcal{F}$ is \textit{contained} within that of $\mathcal{F}_c$.
However, the value $pc = 0$ indicates that $\mathcal{F}$ is invoked by $T$, and not due to the invocation of any other method in $\mathcal{C}$. 
\looseness=-1
\vspace{-2mm}
\begin{definition}
	(\textbf{Schedule}).
	A \textit{schedule} $\mathcal{H} = [e_1, e_2, ..., e_n]$, $\forall e \in \mathcal{H}, e.f \in \{\mathcal{F}.\mathsf{name} |\mathcal{F} \in \vec{\mathcal{F}}\}$ is a valid sequence of $n$ events that can be executed by the EVM.
	The events, when executed in order on an initial contract state $\Delta$, yield the final state $\Delta^\prime$, \ie, $\Delta \xrightarrow{e_1} \Delta_1 \xrightarrow{e_2} \Delta_2... \xrightarrow{e_n} \Delta^\prime$, which we denote as $\Delta \xrightarrow{\mathcal{H}} \Delta^\prime$.
	The set of all possible schedules is denoted by $\mathbb{H}$.
	\looseness=-1
\end{definition}

\vspace{-4mm}
\begin{definition}
	(\textbf{Equivalent schedules}).
	Two schedules $\mathcal{H}_1$ and $\mathcal{H}_2$, where $|\mathcal{H}_1| = |\mathcal{H}_2|$, are \textit{equivalent}, if $\forall e \in \mathcal{H}_1, \exists e^\prime \in \mathcal{H}_2$ such that 
	$e.f = e^\prime.f \land e.inv = e^\prime.inv$, and
		$\forall e^\prime \in \mathcal{H}_2, \exists e \in \mathcal{H}_1$, such that 
	$e^\prime.f = e.f \land e^\prime.inv = e.inv$.
	We denote it by $\mathcal{H}_1 \equiv \mathcal{H}_2$.
	
	Intuitively, equivalent schedules contain the same set of function invocations.
\end{definition}

\vspace{-4mm}
\begin{definition}
	(\textbf{Transformation function}).
	A transformation function $\mu: \mathbb{H} \rightarrow \mathbb{H}$ accepts a schedule $\mathcal{H}$, and transforms it to an equivalent schedule $\mathcal{H}^\prime \equiv \mathcal{H}$, by employing one of two possible strategies at a time---\textbf{(i)} mutates $pc$ of an event $\exists e^\prime \in \mathcal{H}^\prime$, such that $e^\prime.pc$ holds a valid non-zero value, 
	\textbf{(ii)} permutes $\mathcal{H}$.
	These strategies correspond to two possible ways of transaction ordering, respectively:
	\textbf{(a)} when a contract performs an external call, it can be leveraged to re-enter the contract through internal transactions,
	\textbf{(b)} the external transactions of a contract can be mined in any arbitrary order.
\end{definition}

\vspace{-3mm}
\begin{definition}
	(\textbf{State inconsistency bug}).
	For a contract instance $\mathcal{C}$, an initial state $\Delta$, and a schedule $\mathcal{H}_1$ where $\forall e \in \mathcal{H}_1, e.pc = 0$, if there exists a schedule $\mathcal{H}_2 = \mu(\mathcal{H}_1)$, where $\Delta \xrightarrow{\mathcal{H}_1} \Delta_1$ and $\Delta \xrightarrow{\mathcal{H}_2} \Delta_2$, then $\mathcal{C}$ is said to have a \si bug, iff $\Delta_1 \neq \Delta_2$.
\end{definition}

\vspace{-4mm}
\begin{definition}
	(\textbf{Reentrancy bug}).
	If a contract $\mathcal{C}$ contains an SI bug due to two schedules $\mathcal{H}_1$ and $\mathcal{H}_2 = \mu(\mathcal{H}_1)$, such that $\exists e \in \mathcal{H}_2 \;(e.pc \neq 0)$ (first transformation strategy), then the contract is said to have a \reentrancy bug.

	In other words, $e.pc \neq 0$ implies that $e.f$ is a reentrant invocation due to an external call in $\mathcal{C}$.
\end{definition}

\vspace{-4mm}
\begin{definition}
	(\textbf{Generalized TOD bug}).
	If a contract $\mathcal{C}$ contains an SI bug due to two schedules $\mathcal{H}_1$ and $\mathcal{H}_2 = \mu(\mathcal{H}_1)$, such that $\mathcal{H}_2$ is a permutation  (second transformation strategy) of $\mathcal{H}_1$, then the contract is said to have a generalized \tod (G-TOD), or event ordering bug (EO)~\cite{ethracer}.
	
	Permutation of events corresponds to the fact that the transactions can be re-ordered due to the inherent non-determinism in the network, \eg, miner's scheduling strategy, gas supplied, \etc{}
	In this work, we limit the detection to only those cases where \ether transfer is affected by \si---which is in line with the previous work~\cite{securify, oyente}.
	We refer to those as TOD bugs.
\end{definition}
	\section{\explorer: Lightweight exploration over SDG}
\label{static_analysis}
This section introduces the storage dependency graph (SDG), a graph abstraction that captures the control and data flow relations between the storage variables and the critical program instructions, \eg, control-flow deciding, and state-changing operations of a \smart.
To detect SI bugs, we then define \haz, which is modeled as queries over the SDG.
\looseness=-1

\subsection{\textbf{Storage dependency graph (SDG)}}
\label{sec:sdg}
In a \smart, the public methods 
are the entry-points which can be called by an attacker.
\toolname builds a \sdg (SDG) $\mathcal{N} = (V, E, \chi)$ that models the execution flow as if it was subverted by an attacker, and how the subverted flow impacts the global state of the contract.
Specifically, the SDG encodes the following information:

\vspace{1mm}
\noindent
\textbf{Nodes.}
A node of an SDG represents either a storage variable, or a statement operating on a storage variable.
If $\mathcal{V}$ be the set of all storage variables of a contract, and $\mathcal{S}$ be the statements operating on $\mathcal{V}$, the set of nodes $V := \{\mathcal{V} \cup \mathcal{S}\}$.

\vspace{1mm}
\noindent
\textbf{Edges.}
An edge of an SDG represents either the data-flow dependency between a storage variable and a statement, or the relative ordering of statements according to the program control-flow.
$\chi(E) \rightarrow \{\texttt{D}, \texttt{W}, \texttt{O}\}$ is a labeling function that maps an edge to one of the three types.
A directed edge $\langle u,v \rangle$ from node $u$ to node $v$ is labeled as
\textbf{(a)} \texttt{D}; if $u \in \mathcal{V}, v \in \mathcal{S}$, and the statement $v$ is data-dependent on the state variable $u$
\textbf{(b)} \texttt{W}; if $u \in \mathcal{S}, v \in \mathcal{V}$, and the state variable $v$ is written by the statement $u$
\textbf{(c)} \texttt{O}; if $u \in \mathcal{S}, v \in \mathcal{S}$, and statement $u$ precedes statement $v$ in the control-flow graph.

We encode the rules for constructing an SDG in Datalog.
First, we introduce the reader to Datalog preliminaries, and then describe the construction rules.

\noindent
\textbf{Datalog preliminaries.}
A Datalog program  consists of a set of \emph{rules} and a set of \emph{facts}. 
Facts simply declare predicates that evaluate to true. For example, 
{\tt parent("Bill", "Mary")} states that Bill is a parent of Mary. Each Datalog rule defines a predicate as a conjunction of other
predicates. For example, the rule: \texttt{ancestor(x, y) :- parent(x, z), ancestor(z, y)}---says that
\verb+ancestor(x, y)+ is true, if both \verb+parent(x, z)+ and
\verb+ancestor(z, y)+ are true. In addition to variables, predicates can 
also contain constants, 
which are surrounded by double quotes, or
``don't cares'',  denoted by underscores.

\begin{figure}[ht]
	\vspace{-6mm}
	\small
    \[\begin{array}{rlll}
      \mathsf{reach}(s_1, s_2) & :- & \text{$s_2$ is reachable from $s_1$}\\
      \mathsf{intermediate}(s_1, s_2, s_3) & :- & \mathsf{reach}(s_1, s_2), \mathsf{reach}(s_2, s_3)\\
      \mathsf{succ}(s_1, s_2) & :- & \text{$s_2$ is the successor of $s_1$}\\
      \mathsf{extcall}(s, cv) & :- & \text{$s$ is an external call}, \\
                                   & & \text{$cv$ is the call value} \\ 
      \mathsf{entry}(s,m) & :- & \text{$s$ is an entry node of method $m$}\\
      \mathsf{exit}(s,m) & :- & \text{$s$ is an exit node of method $m$}\\
      \mathsf{storage}(v) & :- & \text{$v$ is a storage variable}\\
      \mathsf{write}(s, v) & :- & \text{$s$ updates variable $v$}\\
      \mathsf{depend}(s, v) & :- & \text{$s$ is data-flow dependent on $v$}\\
      \mathsf{owner}(s) & :- & \text{only owner executes $s$} \\
    \end{array}\]
    \vspace{-3.1mm}
    \caption{\small Built-in rules for ICFG related predicates.}
    \label{fig:icfg-rule}
    \vspace{-2.5mm}
\end{figure}

\noindent
\textbf{Base ICFG facts.}
The base facts of our inference engine describe the instructions in the application's inter-procedural control-flow graph (ICFG).
In particular, \Fig{fig:icfg-rule} shows the base rules that are derived from a classical ICFG, where $s$, $m$ and $v$ correspond to a statement,  method, and variable respectively.
\toolname uses a standard static taint analysis out-of-the-box to restrict the entries in the $\mathsf{extcall}$ predicate.
Additionally, $\mathsf{owner}(s)$ represents that $s$ can \textit{only} be executed by contract owners, which enables \toolname to model SI attacks precisely.
Refer to \Appen{app:owner_only_statements} for details.

\begin{figure}[ht]
	\small
	\[\begin{array}{rlll}
	\mathsf{sdg}(s_1, v, \mathtt{'W'}) & :- & \mathsf{write}(s_1, v), \mathsf{storage}(v)\\
	\mathsf{sdg}(s_1, v, \mathtt{'D'}) & :- &  \mathsf{depend}(s_1, v), \mathsf{storage}(v)\\
	\mathsf{sdg}(s_1, s_2, \mathtt{'O'}) & :- &  \mathsf{sdg}(s_1, \_, \_), \mathsf{reach}(s_1, s_2), \mathsf{sdg}(s_2, \_, \_), \\
	& & \neg\mathsf{intermediate}(s_1, \_, s_2)\\
	\mathsf{sdg}(s_1, s_2, \mathtt{'O'}) & :- & \mathsf{extcall}(s_1, \_), \mathsf{entry}(s_2,\_)\\
	\mathsf{sdg}(s_4, s_3, \mathtt{'O'}) & :- & \mathsf{extcall}(s_1, \_), \mathsf{entry}(\_,m_0), \\
	& & \mathsf{succ}(s_1, s_3), \mathsf{exit}(s_4, m_0)
	\end{array}\]
	\vspace{-4mm}
	\caption{\small Rules for constructing SDG.}
	\label{fig:sdg-rule}
	\vspace{-5mm}
\end{figure}

\begin{figure}
	\centering
	\includegraphics[width=0.8\linewidth]{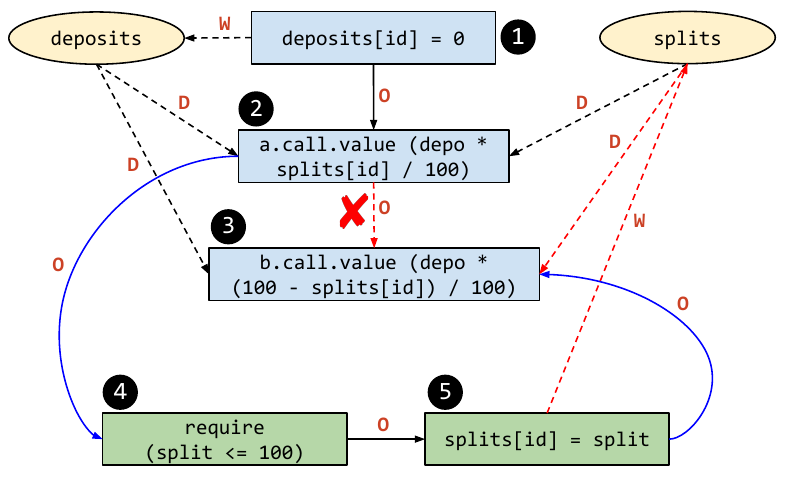}
	\vspace{-2mm}
	\caption{\small SDG for Example 1.
		Ovals and rectangles represent storage variables and instructions.
		Blue [\crule[instrblue]{2.5mm}{2.5mm}] and green [\crule[instrgreen]{2.5mm}{2.5mm}] colored nodes correspond to instructions from \texttt{splitFunds} and \texttt{updateSplit} methods, respectively.
		The \texttt{O}, \texttt{D}, and \texttt{W} edges stand for \texttt{order}, \texttt{data}, and \texttt{write} edges, respectively.
		The red [\crule[edgered]{2.5mm}{2.5mm}] edges on \texttt{splits} denote \haz.}
	\label{fig:sdg-composed}
	\vspace{-6.5mm}
\end{figure}

\noindent
\textbf{SDG construction.}
The basic facts generated from the previous step can be leveraged to construct the SDG.
As shown in Fig~\ref{fig:sdg-rule}, a ``write-edge'' of an SDG is labeled as \texttt{'W'}, and is constructed by checking whether storage variable $v$ gets updated in statement $s$.
Similarly, a ``data-dependency edge'' is labeled as \texttt{'D'}, and is constructed by determining whether the statement $s$ is data-dependent on the storage variable $v$.
Furthermore, we also have the ``order-edge'' to denote the order between two statements, and those edges can be drawn by checking the reachability between nodes in the original ICFG.
Finally, an external call in \solidity can be weaponized by the attacker by hijacking the current execution.
In particular, once an external call is invoked, it may trigger the callback function of the attacker who can perform arbitrary operations to manipulate the storage states of the original contract.
To model these semantics, we also add extra \texttt{'O'}-edges to connect external calls with other public functions that can potentially update storage variables that may influence the execution of the current external call.
Specifically, we add an extra order-edge to connect the external call to the entry point of another public function $m$, as well as an order-edge from the exit node of $m$ to the successor of the original external call.
\vspace{-1.5mm}
\noindent
\begin{example}
Consider Example 1 (\Fig{fig:false_negative}) that demonstrates an SI vulnerability due to both \texttt{splitFunds} and \texttt{updateSplit} methods operating on a state variable \texttt{splits[id]}.
\Fig{fig:sdg-composed} models this attack semantics.
\texttt{deposits} and \texttt{splits[id]} correspond to the variable nodes in the graph.
\Line{12} writes to \texttt{deposits}; thus establishing a \texttt{W} relation from the instruction to the variable node.
\Line{16} and \Line{19} are data-dependent on both the state variables.
Hence, we connect the related nodes with \texttt{D} edges.
Finally, the instruction nodes are linked together with directed \texttt{O} edges following the control-flow.
To model the \reentrancy attack, we created an edge from the external call node \circledtext{2} $\rightarrow$ \circledtext{4}, the entry point of \texttt{splitFunds}.
Next, we remove the edge between the external call \circledtext{2}, and its successor \circledtext{3}.
Lastly, we add an edge between \circledtext{5}, the exit node of \texttt{updateSplit}, and \circledtext{3}, the following instruction in \texttt{updateSplit}.\looseness=-1
\end{example}

\subsection{\textbf{Hazardous access}}
\label{sec:hazardous_access}
Following our discussion in \Sect{si_bugs}, to detect SI bugs in a \smart, one needs to enumerate and evaluate all possible schedules on every contract state---which is computationally infeasible.
To enable scalable detection of SI bugs statically, we define \textit{\haz}, which is inspired by the classical data race problem, where two different execution paths operate on the same storage variable, and at least one operation is a \textit{write}.
In a \smart, the \textit{execution paths} correspond to two executions of public function(s).
\looseness=-1

As shown in the $\mathsf{hazard}(.)$ predicate in \Fig{fig:ce-rule}, a \haz is a tuple denoted by $\langle s_1, s_2, v \rangle$, where $v$ is a storage variable which both the statements $s_1$ and $s_2$ operate on, and either $s_1$, or $s_2$, or both are \textit{write} operations.
While deriving the data-flow dependency predicate $\mathsf{sdg}(s, v, \mathtt{'D'})$, we consider both direct and indirect dependencies of the variable $v$.
We say that a statement $s$ operates on a variable $v$ if either $s$ is an assignment of variable $v$ or $s$ contains an expression that is dependent on variable $s$.

\toolname identifies \haz statically by querying the contract's SDG, which is a path-condition agnostic data structure.
A non-empty query result indicates the existence of a \haz.
However, these accesses might not be feasible in reality due to conflicting path conditions.
The \refiner module (\Sect{refinement}) uses symbolic evaluation to prune such infeasible accesses.

\subsection{\textbf{State inconsistency bug detection}}
\label{sec:si_bug_detection}
As discussed in \Sect{si_bugs}, a \smart{} contains an SI bug if there exists two schedules that result in a different contract state, \ie, the values of the storage variables.
Instead of enumerating all possible schedules (per definition) statically which is computationally infeasible, we use \haz as a \textit{proxy} to detect the root cause of SI.
Two schedules can result in different contract states if:
\textbf{(a)} there exist two operations, where at least one is a \textit{write} access, on a common storage variable, and 
\textbf{(b)} the relative order of such operations differ in two schedules.
The \haz captures the first (\textbf{a}) condition.
Now, in addition to hazardous access, SI bugs require to hold certain conditions that can alter (\textbf{b}) the relative order of the operations in the \haz pair.
For \reentrancy, \toolname checks if a hazardous access pair is reachable in a reentrant execution, as it can alter the execution order of the statements in a hazardous access pair.
To detect TOD, \toolname checks whether an \ether transfer call is reachable from one of the statements in a hazardous access pair.
In this case, the relative execution order of those statements determines the amount of \ether transfer.

\noindent
\textbf{Reentrancy detection.}
A malicious \reentrancy query (\Fig{fig:ce-rule}) 
looks for a hazardous access pair $\langle s_{\mathsf{1}}, s_{\mathsf{2}} \rangle$ such that both $s_{\mathsf{1}}$ and $s_{\mathsf{2}}$ are reachable from an external call in the SDG, and executable by an attacker.

To detect \textit{delegate-based} \reentrancy attacks, where the $\mathsf{delegatecall}$ destination is tainted, we treat $\mathsf{delegatecall}$ in the same way as the  $\mathsf{extcall}$ in \Fig{fig:ce-rule}.
For untainted $\mathsf{delegatecall}$ destinations, if the source code of the delegated contract is available, \toolname constructs an SDG that combines both the contracts.
If neither the source, nor the address of the delegated contract is available, \toolname treats $\mathsf{delegatecall}$ in the same way as an unsafe external call.
For \textit{create-based} attacks, since the source code of the child contract is a part of the parent contract, \toolname builds the SDG by combining both the creator (parent) and the created (child) contracts.
Subsequently, \toolname leverages the existing queries in \Fig{fig:ce-rule} on the combined SDG.
For untainted $\mathsf{extcall}$, and $\mathsf{delegatecall}$ destinations, \toolname performs inter-contract (\Appen{app:inter_contract_analysis}) analysis to build an SDG combining both contracts.
\looseness=-1

\vspace{-2mm}
\noindent
\begin{example}
When run on the SDG in \Fig{fig:sdg-composed} (Example 1), the query returns the tuple $\langle 3, 5 \rangle$, because they both operate on the state variable \texttt{splits}, and belong to distinct public methods, \viz, \texttt{splitFunds} and \texttt{updateSplit} respectively.
\end{example}

\begin{figure}[t]
	\vspace{-2mm}
	\small
	\[\begin{array}{rlll}
	\mathsf{hazard}(s_1, s_2, v) & :- & \mathsf{storage}(v), \mathsf{sdg}(s_{\mathsf{1}}, v, \mathtt{'W'}),\\
	& &\mathsf{sdg}(s_{\mathsf{2}}, v, \_), s_1\neq s_2 \\
	
	\mathsf{reentry}(s_{\mathsf{1}}, s_{\mathsf{2}}) & :- & \mathsf{extcall}(e, \_), \mathsf{reach}(e, s_1), \mathsf{reach}(e, s_2),  \\
	& & \mathsf{hazard}(s_1, s_2, \_), \neg \mathsf{owner}(s_1), \neg \mathsf{owner}(s_2)\\
	
	\mathsf{tod}(s_{\mathsf{1}}, s_{\mathsf{2}}) & :- & \mathsf{extcall}(e, cv), cv > 0, \mathsf{reach}(s_1, e),\\ 
	& & \mathsf{hazard}(s_1, s_2, \_), \neg \mathsf{owner}(s^*),\\ 
	& & s^{\star}\in \{s_1, s_2\} \\
	
	\mathsf{Base \ case:} \\
	\mathsf{cex}(s_0, s_1) & :- & \mathsf{entry}(s_0, \_), \mathsf{succ}(s_0, s_1), \mathsf{f}(s_1,s_2), \\ 
	& &\mathsf{extcall}(s',\_), \mathsf{reach}(s_1, s^{\star}), \\
	& & s^{\star}\in \{s_1,s_2,s'\}, f \in \{\mathsf{tod,reentry}\}\\
	\mathsf{Inductive \ case:} \\
	\mathsf{cex}(s_1, s_2) & :- & \mathsf{cex}(\_, s_1), \mathsf{succ}(s_1, s_2), \mathsf{f}(s_3,s_4), \\ 
	& &\mathsf{extcall}(s',\_),\mathsf{reach}(s_2, s^{\star}),\\ 
	& & s^{\star} \in \{s_3,s_4,s'\}, f \in \{\mathsf{tod,reentry}\}\\
	\end{array}\]
	\vspace{-4mm}
	\caption{\small Rules for hazardous access and counter-examples.}
	\label{fig:ce-rule}
	\vspace{-6mm}
\end{figure}

\vspace{-2mm}
\noindent
\textbf{TOD detection.}
As explained in \Sect{background}, TOD happens when \ether transfer is affected by re-ordering transactions.
Hence, a hazardous pair $\langle s_{\mathsf{1}}, s_{\mathsf{2}} \rangle$ forms a TOD if the following conditions hold: 1) an external call is reachable from either $s_1$ or $s_2$, and 2) the amount of \ether sent by the external call is greater than zero.

\toolname supports all three TOD patterns supported by \securify~\cite{securify}---\textbf{(i)} \texttt{TOD Transfer} specifies that the pre-condition of an \ether transfer, \eg, a condition $c$ guarding the transfer, is influenced by transaction ordering,
\textbf{(ii)} \texttt{TOD Amount} indicates that the amount $a$ of \ether transfer is dependent on transaction ordering, and
\textbf{(iii)} \texttt{TOD Receiver} defines that the external call destination $e$ is influenced by the transaction ordering.
To detect these attacks, \toolname reasons if $c$, or $a$, or $e$ is data-flow dependent on some $\mathsf{storage}(v)$, and the statements corresponding to those three are involved in forming a hazardous pair.\looseness=-1

\noindent
\textbf{Counter-example generation.}
If a %
query over the SDG returns $\bot$ (empty), then the contract is safe, because the SDG models the state inconsistency in the contract. 
On the other hand, if the query returns a list of pairs $\langle s_1, s_2 \rangle$, \toolname performs a \emph{refinement} step to determine if those pairs are indeed feasible.
Since the original output pairs (\ie, $\langle s_1, s_2 \rangle$) can not be directly consumed by the symbolic execution engine, \toolname leverages the \texttt{cex}-rule in Figure~\ref{fig:ce-rule} to compute the minimum ICFG $G$ that contains statements $s_1$, $s_2$, and the relevant external call $s'$. 
In the base case, \texttt{cex}-rule includes edges between entry points and their successors that can transitively reach $s_1$, $s_2$, or $s'$. In the inductive case, for every node $s_1$ that is already in the graph, we recursively include its successors that can also reach $s_1$, $s_2$, or $s'$.
\looseness=-1

\vspace{-2mm}
\begin{example}
\toolname{} extracts the graph slice starting from the root (not shown in \Fig{fig:sdg-composed}) of the SDG to node \circledtext{5}.
The algorithm extracts the sub-graph $\langle \texttt{root} \rangle \!\!\xrightarrow{*}$ \circledtext{2} $\rightarrow$ \circledtext{4} $\rightarrow$ \circledtext{5} $\rightarrow$ \circledtext{3}, maps all the SDG nodes to the corresponding ICFG nodes, and computes the final path slice which the \refiner{} runs on.
\end{example}

	\section{\refiner: Symbolic evaluation with value summary}
\label{refinement}
As explained in \Sect{static_analysis}, if the \explorer module reports an alarm, then there are two possibilities: either the contract is indeed vulnerable, or the current counter-example (\ie, subgraph generated by the rules in \Fig{fig:ce-rule}) is infeasible.
Thus, \toolname proceeds to refine the subgraph by leveraging symbolic evaluation (\Sect{sec:sym-exe}).
However, as we show later in the evaluation, a naive symbolic evaluation whose storage variables are completely unconstrained will raise several false positives.
To address this challenge, the \refiner module in \toolname leverages a light-weight \emph{value summary analysis} (\Sect{sec:value-sum}) that output the potential symbolic values of each storage variable under different constraints, which will be used as the pre-condition of the symbolic evaluation (\Sect{sec:sym-exe}).

\subsection{\textbf{Value summary analysis (VSA)}}
~\label{sec:value-sum} 
For each storage variable, the goal of value summary analysis (VSA) is to compute
its invariant that holds through the life-cycle of a \smart.
While summary-based analysis has been applied in many different applications before, there is no off-the-shelf VSA for \smart s that we could leverage for the following reasons:
\textbf{(a) Precision.}
A value summary based on abstract interpretation~\cite{Pereira13} that soundly computes the interval for each storage variable scales well, but since it ignores the path conditions under which the interval holds, it may lead to \emph{weaker preconditions} that are not sufficient to prune infeasible paths.
For the example in \Fig{fig:false_positive}, a naive and scalable analysis will ignore the control flows, and conclude that the summary of \texttt{mutex} is $\top$ (either \texttt{true} or \texttt{false}), which will be useless to the following symbolic evaluation, since \texttt{mutex} is unconstrained.
\textbf{(b) Scalability}.
A \textit{path-by-path} summary~\cite{Godefroid07,AnandGT08} that relies on symbolic execution first computes the pre-condition $pre_w$, post-condition $post_w$, and per-path summary $\phi_w = pre_w \land post_w$ for every path $w$.
The overall summary $\phi_f$ of the function $f$ is the disjunction of individual path summaries, \ie, $\phi_f = \lor_w \phi_w$.
We identify the following barriers in adopting this approach out of the box:
\textbf{(i)} \textit{Generation}: The approach is computationally intensive due to well-known path explosion problem.
\textbf{(ii)} \textit{Application}: The summary being the unification of the constraints collected along all the paths, such a summary is complex, which poses a significant challenge to the solver.
In fact, when we evaluated~(\Appen{app:extended_evaluation}) our technique by plugging in a similar path-by-path summary, the analysis timed out for $\FPeval{\v}{round(\clintPathSummaryTimeout/\clintSpeedupDataset*100,2)}\v\%$ of the contracts due to the increased cost of the \refine phase.
\textbf{(iii)} \textit{Usability}: Lastly, such a summary is precise, yet expensive.
Computing a precise summary is beneficial only when it is used sufficient times.
Our aim is to build a usable system that scales well in two dimensions---both to large contracts, and a large number of contracts.
As the dataset is deduplicated, the scope of reusability is narrow. 
Therefore, an expensive summary does not pay off well given our use case.
What we need in \toolname is a summarization technique that has a small resource footprint, yet offers reasonable precision for the specific problem domain, \ie, \smart s.

Therefore, we design a domain-specific VSA (\Fig{fig:rules-value-sum}) to tackle both the challenges:
\textbf{(a)} \textbf{Precision}:
Unlike previous scalable summary techniques that map each variable to an interval whose path conditions are merged,
we compensate for such precision loss at the merge points of the control flows using an idea inspired by symbolic union~\cite{rosette}---our analysis stitches the branch conditions to their corresponding symbolic variables at the merge points.
\textbf{(b)} \textbf{Scalability}:
\textbf{(i)} \textit{Generation}: This design choice, while being more precise, could still suffer from path explosion.
To mitigate this issue, our analysis first starts with a precise abstract domain that captures concrete values and their corresponding path conditions, and then \emph{gradually sacrifices} the precision in the context of statements that are difficult, or expensive to reason about, \eg, loops, return values of external calls, updates over nested data structures, \etc{}
\textbf{(ii)} \textit{Application}:
Lastly, we carefully design the evaluation rules (If-rule in \Fig{fig:rules-value-sum}) that selectively drop path conditions at the confluence points---which leads to simpler constraints at the cost of potential precision loss.
However, our evaluation of \toolname suggests that, indeed, our design of VSA strikes a reasonable trade-off in the precision-scalability spectrum in terms of both bug detection and analysis time.

\begin{figure}[!t]
\small
\[
\begin{array}{rcl}
    \text{Program }    \contract  &::=& (\valEnv, \pathEnv, \vec{\func}) \\
    \text{ValueEnv }  \valEnv &::=& V \to \text{Expr} \\
    \text{PathEnv }    \pathEnv   &::=& \mathsf{loc} \to C \\
    \text{Expr }    e &::=& x ~|~ c ~|~ op(\vec{e}) ~|~ S(\vec{e}) \\
    \text{Statement }  s &::=& \mathsf{havoc}(s) ~|~ l := e ~|~ s; s  ~|~ r=f(\vec{e}) \\
    &|  & (\mathtt{if} \ e \ s \ s) ~|~ (\mathtt{while} \ e \ s) \\
    \text{Function }   \func &::=& {\bf function} ~ \emph{f}(\vec{x}) ~ s ~ {\bf returns} ~ y \\   
\end{array}
\]
\[
\begin{array}{c}
x, y \in \textbf{Variable} \quad c \in \textbf{Constant} \quad S \in \textbf{StructName} \\
\end{array}
\]
\vspace{-7mm}
\caption{\small Syntax of our simplified language.}
\label{fig:syntax-ir}
\vspace{-8mm}
\end{figure}

To formalize our rules for VSA, we introduce a simplified language in \Fig{fig:syntax-ir}.
In particular, a contract $\contract$ consists of 
\textbf{(a)} a list of public functions $\vec{\func}$ (private functions are inline), 
\textbf{(b)} a value environment $\valEnv$ that maps variables or program identifiers to concrete or symbolic values, and 
\textbf{(c)} a path environment $\pathEnv$ that maps a location $loc$ to its path constraint $C$.
It is a boolean value encoding the branch decisions taken to reach the current state.
Moreover, each function $\func$ consists of arguments, return values, and a list of statements containing loops, branches, and sequential statements, \etc{}
Our expressions $e$ include common features in \solidity{} such as storage access, struct initialization, 
and arithmetic expressions (function invocation is handled within a statement), \etc{}
Furthermore, since all private functions are inline, we assume that the syntax for calling an external function with return variable $r$ is $r = f(\vec{e})$.
Finally, we introduce a \textsf{havoc} operator to make those variables in hard-to-analyze statements unconstrained, \eg, \textsf{havoc}($s$) changes each variable in $s$ to $\top$ (completely unconstrained). 

\Fig{fig:rules-value-sum} shows a representative subset of the inference rules for computing the summary.
A program state consists of the value environment $\valEnv$ and the path condition $\pathEnv$.
A rule $\langle e, \valEnv,\pathEnv \rangle {\leadsto} \langle v, \valEnv', \pathEnv' \rangle$ says that a successful execution of $e$ in the program state $\langle \valEnv,\pathEnv \rangle$ results in value $v$ and the state $\langle \valEnv',\pathEnv' \rangle$.

\noindent
\textbf{Bootstrapping.}
The value summary procedure starts with the ``contract'' rule that sequentially generates the value summary for each public function $\func_i$ (all non-public methods are inline).
The output value environment $\valEnv'$ contains the value summary for all storage variables.
More precisely, for each storage variable $s$, $\valEnv'$ maps it to a set of pairs $\langle \pathEnv, v \rangle$ where $v$ is the value of $s$ under the constraint $\pathEnv$.
Similarly, to generate the value summary for each function $\func_i$, \toolname applies the ``Func'' rule to visit every statement $s_i$ inside method $\func_i$. 

\noindent
\textbf{Expression.}
There are several rules to compute the rules for different expressions $e$. In particular, if $e$ is a constant $c$, the value summary for $e$ is $c$ itself. 
If $e$ is an argument of a public function $\func_i$ whose values are completely under the control of an attacker, the ``Argument'' rule will havoc $e$ and assume that its value can be any value of a particular type. 

\noindent
\textbf{Helper functions.}
The \texttt{dom}$(\valEnv)$ returns all the keys of an environment $\valEnv$. The \texttt{lhs}$(e)$ returns variables written by $e$. 

\noindent
\textbf{Collections.}
For a variable of type Array or Map, our value summary rules do not differentiate elements under different indices or keys. In particular, for a variable $a$ of type array, the ``store'' rule performs a weak update by unioning all the previous values stored in $a$ with the new value $e_0$. We omit the rule for the map since it is similar to an array. Though the rule is imprecise as it loses track of the values under different indices, it summarizes possible values that are stored in $a$.

\noindent
\textbf{Assignment.}
The ``assign'' rule essentially keeps the value summaries for all variables from the old value environment $\valEnv$ except for mapping $e_0$ to its new value $e_1$.

\noindent
\textbf{External calls.} 
Since all private and internal functions are assumed to be inline, we assume all function invocations are external. 
As we do not know how the attacker is going to interact with the contract via external calls, we assume that it can return arbitrary values. Here is the key intuition of the ``ext" rule: for any invocation to an external function, 
we havoc its return variable $r$.\looseness=-1

\noindent
\textbf{Loop.}
Finally, since computing value summaries for variables inside loop bodies are very expensive and hard to scale to complex contracts, our ``loop'' rule simply havocs all variables that are written in the loop bodies.

\noindent
\textbf{Conditional.}
Rule ``if'' employs a meta-function $\union$ to merge states from alternative execution paths.
\vspace{-1.5mm}
\[
\small
\union(b, v_1, v_2) = 
\begin{cases}
  \{ \langle \top, v_1 \rangle \} & \text{if } b== \mathtt{true}\\    
  \{ \langle \top, v_2 \rangle \} & \text{if } b== \mathtt{false}\\    
  \{ \langle b, v_1 \rangle, \langle \neg b, v_2 \rangle \} & \text{Otherwise }\\     
\end{cases}
\]
In particular, the rule first computes the symbolic expression $v_0$ for the branch condition $e_0$. If $v_0$ is evaluated to \texttt{true}, then the rule continues with the \texttt{then} branch $e_1$ and computes its value summary $v_1$.
Otherwise, the rule goes with the \texttt{else} branch $e_2$ and obtains its value summary $v_2$. Finally, if the branch condition $e_0$ is a symbolic variable whose concrete value cannot be determined, then our value summary will include both $v_1$ and $v_2$ together with their path conditions.
Note that in all cases, the path environment $\pathEnv'$ needs to be computed by conjoining the original $\pathEnv$ with the corresponding path conditions that are taken by different branches.

\begin{figure}[t]
\vspace{-1mm}
\footnotesize
\[
\hspace{-20pt}
\begin{array}{c}

\irulelabel
{\begin{array}{c}
\contract= (\valEnv, \pathEnv, \vec{\func}), \ 
{\langle \func_0, \valEnv,\pathEnv \rangle {\leadsto} \langle \mathtt{void}, \valEnv_1, \pathEnv_1 \rangle}\\
...\\
{\langle \func_n, \valEnv_n,\pathEnv_n \rangle {\leadsto} \langle \mathtt{void}, \valEnv', \pathEnv' \rangle}
\end{array}}
{\langle \contract, \valEnv,\pathEnv \rangle {\leadsto} \langle \mathtt{void}, \valEnv', \pathEnv' \rangle}
{\textrm{(Contract)}} \\ \ \\

\irulelabel
{\begin{array}{c}
    {\langle \stmt, \valEnv,\pathEnv \rangle {\leadsto} \langle \mathtt{void}, \valEnv', \pathEnv' \rangle} \\
\end{array}}
{\langle (\mathtt{function} \ f(\vec{x}) \ s \ \mathtt{returns} \ y), \valEnv,\pathEnv \rangle {\leadsto} \langle \mathtt{void}, \valEnv', \pathEnv' \rangle}
{\textrm{(Func)}} \\ \ \\

\irulelabel
{}
{\langle c, \valEnv,\pathEnv \rangle {\leadsto} \langle c, \valEnv, \pathEnv \rangle}
{\textrm{(Const)}} 
~
\irulelabel
{\mathtt{isArgument}(a) \ v=\mathtt{havoc}(a)}
{\langle a, \valEnv,\pathEnv \rangle {\leadsto} \langle v, \valEnv', \pathEnv \rangle}
{\textrm{(Argument)}} \\ \ \\

\irulelabel
{\begin{array}{c}
    {\langle e_1, \valEnv,\pathEnv \rangle {\leadsto} \langle v_1, \valEnv, \pathEnv \rangle} \quad  \oplus \in \{+,-,*,/ \} \\ 
    {\langle e_2, \valEnv,\pathEnv \rangle {\leadsto} \langle v_2, \valEnv, \pathEnv \rangle} \quad v = v_1 \oplus v_2 \\
\end{array}}
{\langle (e_1 \oplus e_2), \valEnv,\pathEnv \rangle {\leadsto} \langle v, \valEnv, \pathEnv \rangle}
{\textrm{(Binop)}} \\ \ \\

\irulelabel
{\begin{array}{c}
{\langle e_0, \valEnv,\pathEnv \rangle {\leadsto} \langle v_0, \valEnv, \pathEnv \rangle} \\
    \valEnv' = \{ y \mapsto \valEnv(y) \ | \ y \in \mathtt{dom}(\valEnv) \land y \neq a\} \ \cup \ \{a[0] \mapsto (\valEnv(a[0]) \cup \langle \pathEnv, v_0 \rangle) \} \\
\end{array}}
{\langle (a[i] = e_0), \valEnv,\pathEnv \rangle {\leadsto} \langle \mathtt{void}, \valEnv', \pathEnv \rangle \hspace{20pt minus 1fill}\textrm{(Store)}}{} \\ \ \\

\irulelabel
{\begin{array}{c}
    \langle \_, v \rangle = \valEnv(a[0])\\
\end{array}}
{\langle a[i], \valEnv,\pathEnv \rangle {\leadsto} \langle v, \valEnv, \pathEnv \rangle}
{\textrm{(Load)}} \\ \ \\

\irulelabel
{\begin{array}{c}
    \valEnv' = \{ y \mapsto \valEnv(y) \ | \ y \in \mathtt{dom}(\valEnv) \land y \neq e_0\} \ \cup \ \{e_0 \mapsto \langle \pathEnv,e_1 \rangle \cup \valEnv(e_0)\} \\
\end{array}}
{\langle (e_0=e_1), \valEnv,\pathEnv \rangle {\leadsto} \langle \mathtt{void}, \valEnv', \pathEnv \rangle \hspace{20pt minus 1fill}\textrm{(Assign)}}
{} \\ \ \\

\irulelabel
{\begin{array}{c}
    \valEnv' = \{ y \mapsto \valEnv(y) \ | \ y \in \mathtt{dom}(\valEnv) \land y \neq r\} \ \cup \ \{r \mapsto \langle \pathEnv,\mathtt{havoc}(r) \rangle \} \\
\end{array}}
{\langle r=\mathtt{f}(\vec{e}), \valEnv,\pathEnv \rangle {\leadsto} \langle \mathtt{void}, \valEnv', \pathEnv \rangle \hspace{20pt minus 1fill}\textrm{(Ext)}}
{} \\ \ \\

\irulelabel
{\begin{array}{c}
	\langle e_0, \valEnv,\pathEnv \rangle {\leadsto} \langle v_0, \valEnv, \pathEnv \rangle  \quad  \pathEnv' = \pathEnv \land v_0 \\
    \valEnv' = \{ y \mapsto \valEnv(y) \ | \ y \not\in \mathtt{lhs}(e_1)\} \ \cup \ \{y \mapsto \langle \pathEnv',\mathtt{havoc}(y)\rangle \ | \ y \in \mathtt{lhs}(e_1) \} \\
\end{array}}
{\langle (\mathtt{while} \ e_0 \ e_1), \valEnv,\pathEnv \rangle {\leadsto} \langle v_0, \valEnv', \pathEnv \land \neg v_0 \rangle \hspace{20pt minus 1fill}\textrm{(Loop)}}
{} \\ \ \\

\irulelabel
{\begin{array}{c}
{\langle e_0, \valEnv,\pathEnv \rangle {\leadsto} \langle v_0, \valEnv, \pathEnv \rangle} \quad b=isTrue(v_0) \\
{\langle e_1, \valEnv,\pathEnv\land b \rangle {\leadsto} \langle v_1, \valEnv_1, \pathEnv_1 \rangle} \\
{\langle e_2, \valEnv,\pathEnv\land \neg b \rangle {\leadsto} \langle v_2, \valEnv_2, \pathEnv_2 \rangle} \\
{\valEnv' =  \valEnv \ {\cup} \ \valEnv_1 \ {\cup} \valEnv_2 } \\
\end{array}}
{\langle (\mathtt{if} \ e_0 \ e_1 \ e_2), \valEnv,\pathEnv \rangle {\leadsto} \langle \union(b, v_1,v_2), \valEnv', \pathEnv \rangle}
{\textrm{(If)}} \\ \ \\

\end{array}
\]
\vspace{-18pt}
\caption{\small Inference rules for value summary analysis.}
\label{fig:rules-value-sum}
\vspace{-7mm}
\end{figure}

\subsection{\textbf{Symbolic evaluation}}
~\label{sec:sym-exe}
Based on the rules in \Fig{fig:ce-rule}, if the contract contains a pair of statements $\langle s_1, s_2 \rangle$ that match our \si query (\eg, \reentrancy), the \explorer module (\Sect{static_analysis}) returns a subgraph $G$ (of the original ICFG) that contains statement $s_1$ and $s_2$. 
In that sense, checking whether the contract indeed contains the \si bug boils down to a standard reachability problem in $G$: does there exist a valid path $\pi$ that satisfies the following conditions: 1) $\pi$ starts from an entry point $v_0$ of a public method, and 2) following $\pi$ will visit $s_1$ and $s_2$, sequentially.~\footnote{Since TOD transfer requires reasoning about two different executions of the same code, we adjust the goal of symbolic execution for TOD as the following: Symbolic evaluate subgraph $G$ twice (one uses \textit{true} as pre-condition and another uses value summary). The amount of \ether in the external call are denoted as $a_1$, $a_2$, respectively. We report a TOD if $a_1 \neq a_2$. }
Due to the over-approximated nature of our SDG that ignores all path conditions, a valid path in SDG does not always map to a \emph{feasible execution path} in the original ICFG. As a result, we have to symbolically evaluate $G$ and confirm whether $\pi$ is indeed feasible.

A naive symbolic evaluation strategy is to evaluate $G$ by precisely following its control flows while assuming that all storage variables are completely unconstrained ($\top$).
With this assumption, as our ablation study shows (\Fig{fig:chart-ablation}), \toolname fails to refute a significant amount of false alarms.
So, the key question that we need to address is: How can we symbolically check the reachability of $G$ while constraining the range of storage variables without losing too much precision?
This is where VSA comes into play.
Recall that the output of our VSA maps each storage variable into a set of abstract values together with their corresponding path constraints in which the values hold. Before invoking the symbolic evaluation engine, we union those value summaries into a global pre-condition that is enforced through the whole symbolic evaluation. 
\vspace{-2mm}
\begin{example}
Recall in Fig~\ref{fig:false_positive}, the \explorer reports a false alarm due to the over-approximation of the SDG.
We now illustrate how to leverage VSA to refute this false alarm.

\noindent
\textbf{Step 1:} By applying the VSA rules in ~\Fig{fig:rules-value-sum} to the contract in \Fig{fig:false_positive}, \toolname generates the summary for storage variable \texttt{mutex}: $\{\langle \mathsf{mutex=false}, \mathtt{false} \rangle$, $\langle \mathsf{mutex=false}, \mathtt{true} \rangle\}$.
In other words, after invoking any sequence of public functions, %
\texttt{mutex} can be updated to \texttt{true} or \texttt{false}, if pre-condition \texttt{mutex==false} holds.
Here, we omit the summary of other storage variables (\eg, \texttt{userBalance}) for simplicity.

\noindent
\textbf{Step 2:} Now, by applying the symbolic checker 
on the \texttt{withdrawBalance} function for the first time, \toolname generates the following path condition $\pathEnv$: $\mathtt{mutex == false} \  \land \ \mathtt{ userBalance[msg.sender] } > \mathtt{ amount}$
as well as the following program state $\valEnv$ before invoking the external call at \Line{9}: $\valEnv=\{\mathtt{mutex}\mapsto \mathtt{true},...\}$

\noindent
\textbf{Step 3:} After Step 2, the current program state $\valEnv$ indicates that the value of \texttt{mutex} is \texttt{true}.
Note that to execute the \texttt{then-branch} of \texttt{withdrawBalance}, \texttt{mutex} must be \texttt{false}.
Based on the value summary of \texttt{mutex} in Step 1, the pre-condition to set \texttt{mutex} to \texttt{false} is $\mathtt{mutex}=\mathtt{false}$.
However, the pre-condition is \textit{not} satisfiable under the current state $\delta$.
Therefore, although the attacker can re-enter the \texttt{withdrawBalance} method through the callback mechanism, it is impossible for the attacker to re-enter the \texttt{then-branch} at \Line{6}, and trigger the external call at \Line{9}.
Thus, \toolname discards the \reentrancy report %
as false positive.\looseness=-1

\end{example}

	\section{Implementation}
\label{implementation}

\noindent
\textbf{Explorer}. It is a lightweight static analysis that lifts the \smart{} to an SDG. 
The analysis is built on top of the \slither~\cite{slither} framework that lifts \solidity{} source code to its intermediate representation called \textsc{SlithIR}.
\toolname uses \slither's API, including the taint analysis, out of the box.
\looseness=-1

\noindent
\textbf{Refiner}.
\toolname leverages \rosette~\cite{rosette} to symbolically check the feasibility of the counter-examples.
\rosette provides support for symbolic evaluation.
\rosette programs use assertions and symbolic values to formulate queries about program behavior, which are then solved with off-the-shelf SMT solvers.
\toolname uses \texttt{(solve expr)} query that searches for a binding of symbolic variables to concrete values that satisfies the assertions encountered during the symbolic evaluation of the program expression \texttt{expr}.

	\section{Evaluation}
\label{evaluation}
In this section, we describe a series of experiments that are designed to answer the following research questions:
\textbf{RQ1.} How effective is \toolname compared to the existing smart contracts analyzers with respect to vulnerability detection?
\textbf{RQ2.} How scalable is \toolname compared to the existing smart contracts analyzers?
\textbf{RQ3.} How effective is the \refine phase in pruning false alarms?

\subsection{Experimental setup}
\vspace{-1mm}
\noindent
\textbf{Dataset.}
We have crawled the source code of all $\num{\contractsCrawled}$ contracts from Etherscan~\cite{etherscan}, which cover a period until October 31, 2020.
We excluded $\num{\contractsExcluded}$ contracts that either require very old versions ($<$0.3.x) of the \solidity{} compiler, or were developed using the \vyper{} framework. 
As a result, after deduplication, our evaluation dataset consists of $\num{\dataset}$ \solidity{} \smart s.
Further, to gain a better understanding of how each tool scales as the size of the contract increases, we have divided the entire dataset, which we refer to as \textbf{full} dataset, into three mutually-exclusive sub-datasets based on the number of lines of source code---\textbf{small} ($[0, 500)$), \textbf{medium} ($[500, 1000)$), and \textbf{large} ($[1000, \infty)$) datasets consisting of $\num{\smallDataset}$, $\num{\mediumDataset}$, and $\num{\largeDataset}$ contracts, respectively.
We report performance metrics individually for all three datasets.

\noindent
\textbf{Analysis setup.}
We ran our analysis on a Celery v4.4.4~\cite{celery} cluster consisting of six identical machines running Ubuntu 18.04.3 Server, each equipped with Intel(R) Xeon(R) CPU E5-2690 v2@3.00 GHz processor ($40$ core) and $256$ GB memory.

\noindent
\textbf{Analysis of real-world contracts.}
We evaluated \toolname{} against four other static analysis tools, \viz,  \securify~\cite{securify}, \vandal~\cite{vandal}, \mythril~\cite{mythril}, \oyente~\cite{oyente}, and one dynamic analysis tool, \viz, \sereum~\cite{sereum}---capable of finding either \reentrancy, or TOD, or both.
Given the influx of \smart{} related research in recent years, we have carefully chosen a representative subset of the available tools that employ a broad range of minimally overlapping techniques for bug detection.
\smartcheck~\cite{smartcheck} and \slither~\cite{slither} were omitted because their reentrancy detection patterns are identical to \securify's NW (No Write After Ext. Call) signature.\looseness=-1

We run all the static analysis tools, including \toolname, on the full dataset under the analysis configuration detailed earlier.
If a tool supports both \reentrancy and TOD bug types, it was configured to detect both.
We summarize the results of the analyses in \Tbl{tbl:bug_finding}.
For each of the analysis tools and analyzed contracts, we record one of the four possible outcomes--
\textbf{(a)} \textit{safe}: no vulnerability was detected 
\textbf{(b)} \textit{unsafe}: a potential \si bug was detected
\textbf{(c)} \textit{timeout}: the analysis failed to converge within the time budget ($\timeBudget$ minutes)
\textbf{(d)} \textit{error}: the analysis aborted due to infrastructure issues, \eg, unsupported \solidity{} version, or a framework bug, \etc{}
For example, the latest \solidity{} version at the time of writing is $0.8.3$, while \oyente{} supports only up to version $0.4.19$.

\subsection{Vulnerability detection}
\label{sec:vulnerability_detection}
In this section, we report the fraction ($\%$) of \emph{safe}, \emph{unsafe} (warnings), and timed-out contracts reported by each tool with respect to the total number of contracts successfully analyzed by that tool, excluding the ``error'' cases.

\noindent
\textbf{Comparison against other tools.}
\securify, \mythril, \oyente, \vandal, and \toolname{} report potential \reentrancy in $\FPeval{\v}{round(\securifyUnsafeDAO/(\securifySafeDAO+\securifyUnsafeDAO+\securifyTimeout)*100,2)}\v\%$, $\FPeval{\v}{round(\mythrilUnsafeDAO/(\mythrilSafeDAO+\mythrilUnsafeDAO+\mythrilTimeout)*100,2)}\v\%$, $\FPeval{\v}{round(\oyenteUnsafeDAO/(\oyenteSafeDAO+\oyenteUnsafeDAO+\oyenteTimeout)*100,2)}\v\%$, $\FPeval{\v}{round(\vandalUnsafeDAO/(\vandalSafeDAO+\vandalUnsafeDAO+\vandalTimeout)*100,2)}\v\%$, and $\FPeval{\v}{round(\clintUnsafeDAO/(\clintSafeDAO+\clintUnsafeDAO+\clintTimeout)*100,2)}\v\%$ of the contracts.
Though all five static analysis tools detect \reentrancy bugs, TOD detection is supported by only three tools, \ie, \securify, \oyente, and \toolname which raise potential TOD warnings in $\FPeval{\v}{round(\securifyUnsafeTOD/(\securifySafeTOD+\securifyUnsafeTOD+\securifyTimeout)*100,2)}\v\%$, $\FPeval{\v}{round(\oyenteUnsafeTOD/(\oyenteSafeTOD+\oyenteUnsafeTOD+\oyenteTimeout)*100,2)}\v\%$, and $\FPeval{\v}{round(\clintUnsafeTOD/(\clintSafeTOD+\clintUnsafeTOD+\clintTimeout)*100,2)}\v\%$ of the contracts.

\begin{table}[t]
	\centering
	\begin{tabular}{l|lrrrr}
		\toprule
		\textbf{Bug} & \textbf{Tool} & \multicolumn{1}{c}{\textbf{Safe}} & \multicolumn{1}{c}{\textbf{Unsafe}} & \multicolumn{1}{c}{\textbf{Timeout}} & \multicolumn{1}{c}{\textbf{Error}} \\ 
		\midrule
		
		\rowcolor{black!10}	\cellcolor{white} & \securify & $\num{\securifySafeDAO}$ & $\num{\securifyUnsafeDAO}$ & $\num{\securifyTimeout}$ & $\num{\securifyError}$ \\
		& \vandal & $\num{\vandalSafeDAO}$ & $\num{\vandalUnsafeDAO}$ & $\num{\vandalTimeout}$ & $\num{\vandalError}$ \\
		\rowcolor{black!10}	\cellcolor{white} & \mythril & $\num{\mythrilSafeDAO}$ & $\num{\mythrilUnsafeDAO}$ & $\num{\mythrilTimeout}$ & $\num{\mythrilError}$ \\
		& \oyente & $\num{\oyenteSafeDAO}$ & $\num{\oyenteUnsafeDAO}$ & $\num{\oyenteTimeout}$ & $\num{\oyenteError}$ \\
		\rowcolor{black!10} \cellcolor{white} \multirow{-5}{*}{\rotatebox{90}{Reentrancy}} & \toolname & $\num{\clintSafeDAO}$ & $\num{\clintUnsafeDAO}$ & $\num{\clintTimeout}$ & $\num{\clintError}$ \\
		\midrule
		
		\rowcolor{black!10}	\cellcolor{white} & \securify & $\num{\securifySafeTOD}$ & $\num{\securifyUnsafeTOD}$ & $\num{\securifyTimeout}$ & $\num{\securifyError}$ \\
		& \oyente & $\num{\oyenteSafeTOD}$ & $\num{\oyenteUnsafeTOD}$ & $\num{\oyenteTimeout}$ & $\num{\oyenteError}$ \\
		\rowcolor{black!10}	\cellcolor{white} \multirow{-3}{*}{\rotatebox{90}{TOD}} & \toolname & $\num{\clintSafeTOD}$ & $\num{\clintUnsafeTOD}$ & $\num{\clintTimeout}$ & $\num{\clintError}$  \\
		\bottomrule
	\end{tabular}
	\vspace{-1.5mm}
	\caption{\small Comparison of bug finding abilities of tools}
	\label{tbl:bug_finding}
	\vspace{-6mm}
\end{table}

\mythril, being a symbolic execution based tool, demonstrates obvious scalability issues: It timed out for $\FPeval{\v}{round(\mythrilTimeout/(\mythrilSafeDAO+\mythrilUnsafeDAO+\mythrilTimeout)*100,2)}\v\%$ of the contracts.
Though \oyente is based on symbolic execution as well, it is difficult to properly assess its scalability.
The reason is that \oyente failed to analyze most of the contracts in our dataset due to the unsupported \solidity version, which explains the low rate of warnings that \oyente emits.
Unlike symbolic execution, static analysis seems to scale well.
\securify{} timed-out for only $\FPeval{\v}{round(\securifyTimeout/(\securifySafeDAO+\securifyUnsafeDAO+\securifyTimeout)*100,2)}\v\%$ of the contracts, which is significantly lower than that of \mythril.
When we investigated the reason for \securify timing out, it appeared that the \texttt{Datalog}-based data-flow analysis (that \securify relies on) fails to reach a fixed-point for larger contracts.
\vandal's static analysis is inexpensive and shows good scalability, but suffers from poor precision. In fact, \vandal{} flags as many as $\FPeval{\v}{round(\vandalUnsafeDAO/(\vandalSafeDAO+\vandalUnsafeDAO+\vandalTimeout)*100,2)}\v\%$ of all contracts as vulnerable to \reentrancy--which makes \vandal{} reports hard to triage due to the overwhelming amount of warnings.
\vandal{} timed out for the least ($\FPeval{\v}{round(\vandalTimeout/(\vandalSafeDAO+\vandalUnsafeDAO+\vandalTimeout)*100,2)}\v\%$) number of contracts.
Interestingly, \securify{} generates fewer \reentrancy warnings than \mythril.
This can be attributed to the fact that the \texttt{NW} policy of \securify{} considers a write after an external call as vulnerable, while \mythril{} conservatively warns about both read and write.
However, \toolname strikes a balance between both scalability and precision as it timed-out only for $\FPeval{\v}{round(\clintTimeout/(\clintSafeDAO+\clintUnsafeDAO+\clintTimeout)*100,2)}\v\%$ of the contracts, and generates the fewest alarms.\looseness=-1

\noindent
\textbf{Ground truth determination.}
\label{subsec:groundtruth}
To be able to provide better insights into the results, we performed manual analysis on a randomly sampled subset of $\manualAnalysisDataset$ contracts ranging up to $3,000$ lines of code, out of a total of $\num{\allToolsSuccess}$ contracts successfully analyzed by all five static analysis tools, without any timeout or error.
We believe that the size of the dataset is in line with prior work~\cite{ethor,zeus}.
We prepared the ground truth by manually inspecting the contracts for \reentrancy and TOD bugs using the %
following criteria:
\textbf{(a)} \textit{Reentrancy:} The untrusted external call allows the attacker to re-enter the contract, which makes it possible to operate on an inconsistent internal state.
\textbf{(b)} \textit{TOD:} A front-running transaction can divert the control-flow, or alter the \ether-flow, \eg, \ether amount, call destination, \etc, of a previously scheduled transaction.

In the end, the manual analysis identified $\groundTruthDAO$ and $\groundTruthTOD$ contracts with \reentrancy and TOD vulnerabilities, respectively.
We then ran each tool on this dataset, and report the number of correct (TP), incorrect (FP), and missed (FN) detection by each tool in \Tbl{tbl:ground_truth}.
For both \reentrancy and TOD, \toolname detected all the vulnerabilities (TP) with zero missed detection (FN), while maintaining the lowest false positive (FP) rate.
We discuss the FPs and FNs of the tools in the subsequent sections.\looseness=-1

\begin{table}[t]
	\centering
	\scriptsize
	\begin{tabular}{l|rrr|rrr}
		\toprule
		\multirow{2}{*}{\textbf{Tool}} & \multicolumn{3}{c|}{\textbf{Reentrancy}} & \multicolumn{3}{c}{\textbf{TOD}} \\ 
		\cline{2-7} 
		& \multicolumn{1}{c}{\textbf{TP}}& \textbf{FP} & \multicolumn{1}{c|}{\textbf{FN}} 
		& \multicolumn{1}{c}{\textbf{TP}} & \multicolumn{1}{c}{\textbf{FP}} & \multicolumn{1}{c}{\textbf{FN}} \\ 
		\midrule
		\rowcolor{black!10}	\securify & $\securifyTPDAO$ & $\securifyFPDAO$ & $\securifyFNDAO$ & $\securifyTPTOD$ & $\securifyFPTOD$ & $\securifyFNTOD$ \\
		\vandal & $\vandalTPDAO$ & $\vandalFPDAO$ & $\vandalFNDAO$ & \vandalTPTOD & \vandalFPTOD & \vandalFNTOD \\
		\rowcolor{black!10} \mythril & $\mythrilTPDAO$ & $\mythrilFPDAO$ & $\mythrilFNDAO$ & \mythrilTPTOD & \mythrilFPTOD & \mythrilFNTOD \\
		\oyente & $\oyenteTPDAO$ & $\oyenteFPDAO$ & $\oyenteFNDAO$ & $\oyenteTPTOD$ & $\oyenteFPTOD$ & $\oyenteFNTOD$ \\
		\rowcolor{black!10} \toolname & $\clintTPDAO$ & $\clintFPDAO$ & $\clintFNDAO$ & $\clintTPTOD$ & $\clintFPTOD$ & $\clintFNTOD$ \\
		\bottomrule
	\end{tabular}
	\vspace{-1mm}
	\caption{\small Manual determination of the ground truth}
	\label{tbl:ground_truth}
	\vspace{-6.5mm}
\end{table}

\noindent
\textbf{False positive analysis.}
While reasoning about the false positives generated by different tools for the \reentrancy bug, we observe that both \vandal{} and \oyente consider every external call to be reentrant if it can be reached in a recursive call to the calling contract.
However, a reentrant call is \textit{benign} unless it operates on an inconsistent state of the contract.
\securify{} considers \solidity{} \texttt{send} and \texttt{transfer} APIs as external calls, and raisesd violation alerts.
Since the gas limit ($2,300$) for these APIs is inadequate to mount a \reentrancy attack, we refrain from modeling these APIs in our analysis.
Additionally, \securify{} failed to identify whether a function containing the external call is access-protected, \eg, it contains the \texttt{msg.sender == owner} check, which prohibits anyone else but only the contract owner from entering the function.
For both the cases above, though the \explorer detected such functions as potentially unsafe, the benefit of symbolic evaluation became evident as the \refiner{} eliminated these alerts in the subsequent phase.
\mythril{} detects a state variable read after an external call as malicious \reentrancy.
However, if that variable is not written in any other function, that deems the read \textit{safe}.
Since \toolname{} looks for \textit{hazardous access} as a pre-requisite of \reentrancy, it does not raise a warning there.
However, \toolname{} incurs false positives due to imprecise static taint analysis.
A real-world case study of such a false positive is presented in \Appen{sec:case_study_reent_fp}.\looseness=-1

To detect TOD attacks, \securify checks for \emph{writes} to a storage variable that influences an \ether-sending external call.
We observed that several contracts flagged by \securify have storage writes inside the contract's constructor.
Hence, such writes can only happen once during contract creation.
Moreover, several contracts flagged by \securify have both storage variable writes, and the \ether sending external call inside methods which are guarded by predicates like \texttt{require(msg.sender == owner)}---limiting access to these methods only to the contract owner.
Therefore, these methods cannot be leveraged to launch a TOD attack.
\toolname prunes the former case during the \explore phase itself.
For the latter, \toolname leverages the \refine phase, where it finds no difference in the satisfiability of two different symbolic evaluation traces.
In \Appen{sec:case_study_reent_fp}, we present a real-world case  where both \securify and \toolname incur a false positive due to insufficient reasoning of contract semantics.
\looseness=-1

\noindent
\textbf{False negative analysis.}
\securify{} missed valid \reentrancy bugs because it considers only \ether sending call instructions.
In reality, any call can be leveraged to trigger \reentrancy by transferring control to the attacker if its destination is tainted.
To consider this scenario, \toolname{} carries out a taint analysis to determine external calls with tainted destinations.
Additionally, \securify missed \reentrancy bugs due to lack of support for destructive write (DW), and delegate-based patterns.
False negatives incurred by \mythril{} are due to its incomplete state space exploration within specified time-out.
Our manual analysis did not observe any missed detection by \toolname.\looseness=-1

\noindent
\textbf{Finding zero-day bugs using \toolname.}
In order to demonstrate that \toolname is capable of finding zero-day vulnerabilities, we first identified the contracts flagged only by \toolname, but no other tool.
Out of total $\clintOnlyDAO$ \reentrancy-only and $\clintOnlyTOD$ TOD-only contracts, we manually selected $\clintOnlyDAOTriage$ and $\clintOnlyTODTriage$ contracts, respectively.
We limited our selection effort only to contracts that contain at most $500$ lines of code, and are relatively easier to reason about in a reasonable time budget.
Our manual analysis confirms $\zeroDays$ contracts are \textit{exploitable} (not just \textit{vulnerable})---meaning that they can be leveraged by an attacker to accomplish a malicious goal, \eg, draining \ether, or corrupting application-specific metadata, thereby driving the contract to an unintended state.
We present a few vulnerable patterns, and their exploitability in \Appen{sec:case_study_zero_day}.

\noindent
\textbf{Exploitability of the bugs.}
We classified the true alerts emitted by \toolname{} into the following categories---An \textit{exploitable} bug leading to the stealing of \ether, or application-specific metadata corruption (\eg, an index, a counter, \etc), and a \textit{non-exploitable} yet vulnerable bug that can be reached, or triggered (unlike a false positive), but its side-effect is not persistent.
For example, a reentrant call (the attacker) is able to write to some state variable $V$ in an unintended way.
However, along the flow of execution, $V$ is overwritten, and its \textit{correct} value is restored.
Therefore, the effect of \reentrancy did not persist.
Another example would be a state variable that is incorrectly modified during the reentrant call, but the modification does not interfere with the application logic, \eg, it is just written in a log.
Out of the $\zeroDays$ zero-day bugs that \toolname discovered, $11$ allow an attacker to drain Ethers, and for the remaining $36$ contracts, the bugs, at the very least (minimum impact), allow the attacker to corrupt contract metadata---leading to detrimental effects on the underlying application.
For example, during our manual analysis, we encountered a vulnerable contract implementing a housing tracker that the allowed addition, removal, and modification of housing details.
If a house owner adds a new house, the contract mandates the old housing listing to become inactive, \ie, at any point, there can only be one house owned by an owner that can remain in an active state.
However, we could leverage the \reentrancy bug in the contract in a way so that an owner can have more than one active listing.
Therefore, these $36$ contracts could very well be used for stealing Ethers as well, however, we did not spend time and effort to turn those into exploits as this is orthogonal to our current research goal.\looseness=-1

\begin{table}[]
	\centering
	\begin{tabular}{l|rrrr}
		\toprule
		\textbf{Tool} & \multicolumn{1}{c}{\textbf{Small}} & \multicolumn{1}{c}{\textbf{Medium}} & \multicolumn{1}{c}{\textbf{Large}} & \multicolumn{1}{c}{\textbf{Full}} \\
		\midrule 
		\rowcolor{black!10} \securify & $\securifyAnalysisTimeSmall$ & $\securifyAnalysisTimeMedium$ & $\securifyAnalysisTimeLarge$ & $\securifyAnalysisTimeFull$ \\
		\vandal & $\vandalAnalysisTimeSmall$ & $\vandalAnalysisTimeMedium$ & $\vandalAnalysisTimeLarge$ & $\vandalAnalysisTimeFull$ \\
		\rowcolor{black!10} \mythril & $\mythrilAnalysisTimeSmall$ & $\num{\mythrilAnalysisTimeMedium}$ & $\num{\mythrilAnalysisTimeLarge}$ & $\mythrilAnalysisTimeFull$ \\
		\oyente & $\oyenteAnalysisTimeSmall$ & $\oyenteAnalysisTimeMedium$ & $\oyenteAnalysisTimeLarge$ & $\oyenteAnalysisTimeFull$ \\ 
		\rowcolor{black!10} \toolname & $\clintAnalysisTimeSmall$ & $\clintAnalysisTimeMedium$ & $\clintAnalysisTimeLarge$ & $\clintAnalysisTimeFull$ \\
		\bottomrule
	\end{tabular}
	\vspace{-1.5mm}
	\caption{\small Analysis times (in seconds) on four datasets.
	}
	\label{tbl:performance}
	\vspace{-6.5mm}
\end{table}

\noindent
\textbf{Comparison against \textsc{Sereum}.}
Since \sereum is not publicly available, we could only compare \toolname{} on the contracts in their released dataset.
\sereum~\cite{sereum} flagged total $16$ contracts for potential reentrancy attacks, of which $6$ had their sources available in the \etherscan, and therefore, could be analyzed by \toolname.
Four out of those $6$ contracts were developed for old \solidity{} versions ($<$0.3.x)---not supported by our framework.
We ported those contracts to a supported \solidity{} version ($0.4.14$) by making minor syntactic changes not related to their functionality.
According to \sereum, of those $6$ contracts, only one (\texttt{TheDAO}) was a true vulnerability, while five others were its false alarms.
While \toolname{} correctly detects \texttt{TheDAO} as \textit{unsafe}, it raises a false alarm for another contract (\texttt{CCRB}) due to imprecise modeling of untrusted external call.\looseness=-1
\begin{mdframed}[style=graybox]
	\textbf{RQ1}: \toolname emits the fewest warnings in the full dataset, and finds $\zeroDays$ zero-day vulnerabilities.
	On our manual analysis dataset, \toolname detects all the vulnerabilities with the lowest false positive rate.
\end{mdframed}

\subsection{Performance analysis}
\Tbl{tbl:performance} reports the average analysis times for each of the small, medium, and large datasets along with the full dataset.
As the data shows, the analysis time increases with the size of the dataset for all the tools.
\vandal~\cite{vandal} is the fastest analysis across all the four datasets with an average analysis time of $\vandalAnalysisTimeFull$ seconds with highest emitted warnings ($\FPeval{\v}{round(\vandalUnsafeDAO/(\vandalSafeDAO+\vandalUnsafeDAO+\vandalTimeout)*100,2)}\v\%$).
\securify~\cite{securify} is approximately $\FPeval{\v}{round(\securifyAnalysisTimeFull/\clintAnalysisTimeFull,0)}\v$x more expensive than \vandal{} over the entire dataset.
The average analysis time of \mythril~\cite{mythril} is remarkably high ($\mythrilAnalysisTimeFull$ seconds), which correlates with its high number of time-out cases ($\FPeval{\v}{round(\mythrilTimeout/(\mythrilSafeDAO+\mythrilUnsafeDAO+\mythrilTimeout)*100,2)}\v\%$).
In fact, \mythril 's analysis time even for the small dataset is as high as $\mythrilAnalysisTimeSmall$ seconds.
However, another symbolic execution based tool \oyente~\cite{oyente} 
has average analysis time 
close to $\FPeval{\v}{round(\oyenteAnalysisTimeFull/\mythrilAnalysisTimeFull*100,0)}\v\%$ to that of \mythril, as it fails to analyze most of the medium to large contracts due to the unsupported \solidity{} version.
Over the entire dataset, \toolname{} %
 takes as low as $\clintAnalysisTimeFull$ seconds %
with mean analysis times of $\clintAnalysisTimeSmall$, $\clintAnalysisTimeMedium$, and $\clintAnalysisTimeLarge$ seconds for small, medium, and large ones, respectively.
The mean static analysis time is $\clintAverageSATime$ seconds
as compared to the symbolic evaluation phase, which takes $\clintAverageSETime$ seconds.
The value summary computation 
has a mean analysis time of $\clintAverageVSATime$ seconds.\looseness=-1

\begin{mdframed}[style=graybox]
	\textbf{RQ2}: While the analysis time of \toolname is comparable to that of \vandal, it is $\FPeval{\v}{round(\securifyAnalysisTimeFull/\clintAnalysisTimeFull,0)}\v$, $\FPeval{\v}{round(\mythrilAnalysisTimeFull/\clintAnalysisTimeFull,0)}\v$, and $\FPeval{\v}{round(\oyenteAnalysisTimeFull/\clintAnalysisTimeFull,0)}\v$ times faster than \securify, \mythril, and \oyente, respectively.
\end{mdframed}
\vspace{-2mm}

\begin{figure}[!t]
	\vspace{-4mm}
	\centering
	\includegraphics[width=0.8\linewidth]{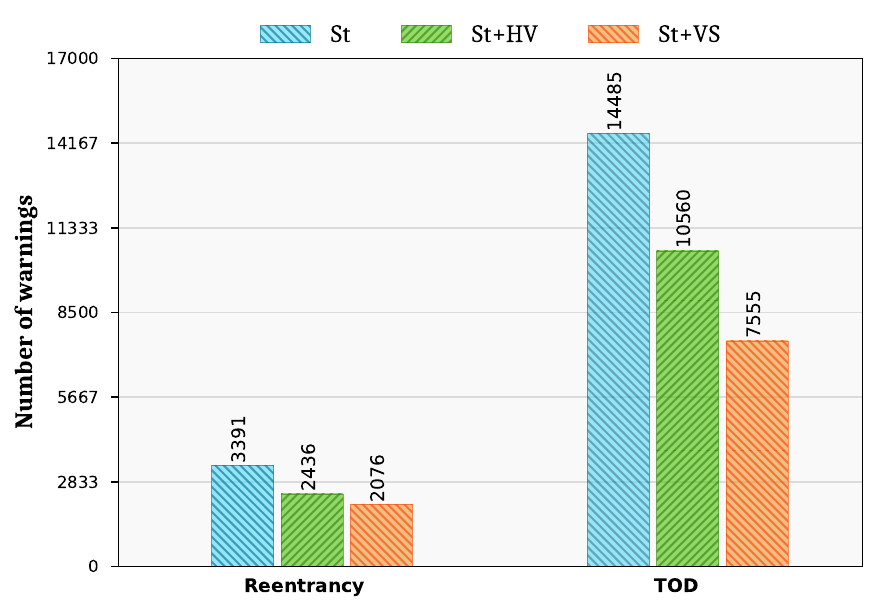}
	\vspace{-3.5mm}
	\caption{\small Ablation study showing the effectiveness of \vsa for \reentrancy and TOD detection.
	}
	\label{fig:chart-ablation}
	\vspace{-7mm}
\end{figure}

\subsection{Ablation study}
\label{sec:ablation}
\noindent
\textbf{Benefit of \vsa:}
To gain a better understanding of the benefits of the symbolic evaluation (\refine) and the \vsa (VSA), we performed an ablation study by configuring \toolname{} in three distinct modes:
\textbf{(a)} \textit{static-only} (\textbf{SO}), only the \explorer runs, and %
\textbf{(b)} \textit{static + havoc} (\textbf{St+HV}), the \refiner{} runs, but it \textit{havocs} all the state variables after the external call. 
\textbf{(c)} \textit{static + value summary} (\textbf{St+VS}), the \refiner{} runs, and it is supplied with the value summary facts that the \explorer computes.
\Fig{fig:chart-ablation} shows the number of warnings emitted by \toolname{} in each of the configurations.
In \textbf{SO} mode, the \explore phase generates $\num{\clintStaticOnlyDAO}$ \reentrancy and $\num{\clintStaticOnlyTOD}$ TOD warnings, which accounts for $\FPeval{\v}{round(\clintStaticOnlyDAO/(\clintSafeDAO+\clintUnsafeDAO+\clintTimeout)*100,2)}\v\%$ and $\FPeval{\v}{round(\clintStaticOnlyTOD/(\clintSafeTOD+\clintUnsafeTOD+\clintTimeout)*100,2)}\v\%$ of the contracts, respectively.
Subsequently, \textbf{St+HV} mode brings down the number of \reentrancy and TOD warnings to $\num{\clintHavocDAO}$ and $\num{\clintHavocTOD}$, which is a $\FPeval{\v}{round((\clintStaticOnlyDAO-\clintHavocDAO)/\clintStaticOnlyDAO*100,2)}\v\%$ and $\FPeval{\v}{round((\clintStaticOnlyTOD-\clintHavocTOD)/\clintStaticOnlyTOD*100,2)}\v\%$ reduction with respect to the \textbf{SO} baseline. 
Lastly, by leveraging value summary, \toolname{} generates $\num{\clintUnsafeDAO}$ \reentrancy and $\num{\clintUnsafeTOD}$ TOD warnings in \textbf{St+VS} mode, which is a $\FPeval{\v}{round((\clintHavocDAO-\clintUnsafeDAO)/\clintHavocDAO*100,2)}\v\%$ and $\FPeval{\v}{round((\clintHavocTOD-\clintUnsafeTOD)/\clintHavocTOD*100,2)}\v\%$ improvement over \textbf{St+HV} configuration.
This experiment demonstrates that our symbolic evaluation and VSA are indeed effective to prune false positives.
\Appen{app:vsa_advantage} presents a real-world case study showing the advantage of VSA.
Additionally, we discuss the relative performance of our VSA over a path-by-path summary technique in \Appen{app:extended_evaluation}.

\begin{mdframed}[style=graybox]
	\textbf{RQ3}: Our symbolic evaluation guided by VSA plays a key role in achieving high precision and scalability.
\end{mdframed}

\section{Limitations}

\noindent
\textbf{Source-code dependency.}
Although \toolname{} is built on top of the \slither~\cite{slither} framework, which requires access to the source code, we do not rely on any rich semantic information from the contract source to aid our analysis.
In fact, our choice of source code was motivated by our intention to build \toolname{} as a tool for developers, while enabling easier debugging and introspection as a side-effect.
Our techniques are not tied to source code, and could be applied directly to bytecode by porting the analysis on top of a contract decompiler that supports variable and CFG recovery.

\noindent
\textbf{Potential unsoundness.}
We do not claim soundness with respect to the detection rules of \reentrancy and TOD bugs.
Also, the meta-language our \vsa is based on distills the core features of the \solidity language, it is not expressive enough to model all the complex aspects~\cite{Jiao20}, \eg, exception propagation, transaction reversion, out-of-gas, \etc{}
In turn, this becomes the source of unsoundness of the \refiner.
Additionally, \toolname{} relies on \slither~\cite{slither} for static analysis.
Therefore, any soundness issues in \slither{}, \eg, incomplete call graph construction due to indirect or unresolved external calls, inline assembly%
, \etc, will be propagated to \toolname.\looseness=-1
	\section{Related work}
\label{related_works}

\noindent
\textbf{Static analysis.}
Static analysis tools such as \securify~\cite{securify}, \madmax~\cite{madmax}, \zeus~\cite{zeus}, \smartcheck~\cite{smartcheck}, and \slither~\cite{slither} %
detect specific vulnerabilities in smart contracts.
Due to their reliance on bug patterns, they over-approximate program states, which can cause false positives and missed detection of bugs.
To mitigate this issue, %
we identified two complementary causes of SI bugs---Stale read and Destructive write.
While the former is more precise than the patterns found in the previous work, the latter, which is not explored in the literature, plays a role in the missed detection of bugs (\Sect{motivation}).
Unlike \toolname, which focuses on SI bugs, \madmax~\cite{madmax} uses a logic-based paradigm to target gas-focused vulnerabilities.
\securify~\cite{securify} first computes control and data-flow facts, and then checks for compliance and violation signatures.
\slither~\cite{slither} uses data-flow analysis to detect bug patterns scoped within a single function.
The bugs identified by these tools are either \textit{local} in nature, or they refrain from doing any path-sensitive reasoning---leading to spurious alarms.
To alleviate this issue, \toolname introduces %
the \refine{} phase that prunes significant numbers of false alarms.\looseness=-1

\noindent
\textbf{Symbolic execution.}
\mythril~\cite{mythril}, \oyente~\cite{oyente}, \ethbmc~\cite{ethbmc}, \smartscopy~\cite{smartscopy}, and \manticore~\cite{manticore} rely on symbolic execution to explore the state-space of the contract.
\ethbmc~\cite{ethbmc}, a bounded model checker, models \texttt{EVM} transactions as state transitions.
\teether~\cite{teether} generates constraints along a critical path having attacker-controlled instructions.
These tools suffer from the limitation of traditional symbolic execution, \eg, path explosion, and do not scale well.
However, \toolname{} uses the symbolic execution \textit{only} for validation, \ie, 
it resorts to under-constrained symbolic execution aided by VSA that over-approximates the preconditions required to update the state variables 
across all executions.\looseness=-1

\noindent
\textbf{Dynamic analysis.}
While \sereum~\cite{sereum} and \soda~\cite{soda} perform run-time checks within the context of a modified \texttt{EVM}, \txspector~\cite{txspector} performs a post-mortem analysis of transactions.
\ecf~\cite{ecf} detects if the execution of a \smart{} is \textit{effectively callback-free} (ECF), \ie, it checks if two execution traces, with and without callbacks, are equivalent---a property that holds for a contract not vulnerable to \reentrancy attacks.
\toolname generalizes %
ECF with the notion of \haz for SI attacks.
Thus, \toolname is not restricted to \reentrancy, instead, can express all properties that are caused by state inconsistencies.
Dynamic analysis tools~\cite{contractfuzzer, harvey, bran, echidna, sfuzz}
rely on manually-written test oracles to detect violations in response to inputs generated according to blackbox or greybox strategies.
Though precise, these tools lack coverage---which is not an issue for static analysis tools, such as \toolname.\looseness=-1

\noindent
\textbf{State inconsistency (SI) notions.}
\serif~\cite{cecchetti21} detects \reentrancy attacks using a notion of trusted-untrusted computation that happens when a low-integrity code, invoked by a high-integrity code, calls back into the high-integrity code before returning.
Code components are explicitly annotated with information flow (trust) labels, which further requires a semantic understanding of the contract.
Then, they design a type system that uses those trust labels to enforce secure information flow through the use of a combination of static and dynamic locks.
However, this notion is unable to capture TOD vulnerabilities, another important class of 
SI bugs.
In \toolname, we take a different approach where we define SI bugs in terms of the side-effect, \ie, creation of an inconsistent state, of a successful attack.
Further, we model the possibility of an inconsistent state resulting from such an attack through \haz.
Perez \etal~\cite{perez21}, \vandal~\cite{vandal}, \oyente~\cite{oyente} consider \reentrancy to be the possibility of being able to re-enter the calling function.
Not only do these tools consider only single-function \reentrancy, but also the notion encompasses legitimate (benign) \reentrancy scenarios~\cite{sereum}, \eg, ones that arise due to withdrawal pattern in \solidity.
In addition, \toolname requires the existence of \haz, which enables us to account for cross-function \reentrancy bugs, as well as model only malicious \reentrancy scenarios.
To detect \reentrancy, \securify~\cite{securify} looks for the violation of the ``no write after external call'' (NW) pattern, which is similar to the ``Stale Read'' (SR) notion of \toolname.
Not all the tools that support \reentrancy bugs have support for TOD.
While \toolname shares its notion of TOD with \securify, \oyente marks a contract vulnerable to TOD if two traces have different \ether flows.
Unlike \toolname for which \haz is a pre-requisite, \oyente raises alarm for independent \ether flows not even related to SI.

	\section{Conclusion}
\label{conclusions}
We propose \toolname, a scalable hybrid tool for automatically identifying SI bugs in smart contracts. \toolname combines lightweight exploration phase followed by symbolic evaluation aided by our novel VSA. 
On the \etherscan dataset, \toolname significantly outperforms  state of the art analyzers in terms of precision, and performance, identifying $47$ previously unknown vulnerable (and exploitable) contracts.

	\section{Acknowledgments}
\label{sec:acknowledgement}
We want to thank our anonymous shepherd and anonymous reviewers for their valuable comments and feedback to improve our paper.
This research is supported by DARPA under the agreement number HR001118C006, by the NSF under awards CNS-1704253, and 1908494, by the ONR under award N00014-17-1-2897, and by the Google Faculty Research Award.
The U.S. Government is authorized to reproduce and distribute reprints for Governmental purposes notwithstanding any copyright notation thereon.
The views and conclusions contained herein are those of the authors
and should not be interpreted as necessarily representing the official policies or endorsements, either expressed or implied, of DARPA or the U.S. Government.
	
	\bibliographystyle{plain}
	\bibliography{bibs/main} 
	\appendices

\section{Extended Evaluation}
\label{app:extended_evaluation}
\begin{figure}[!t]
	\centering
	\setlength{\fboxsep}{0pt}%
	\setlength{\fboxrule}{0.5pt}%
	\fbox{\includegraphics[width=0.8\linewidth]{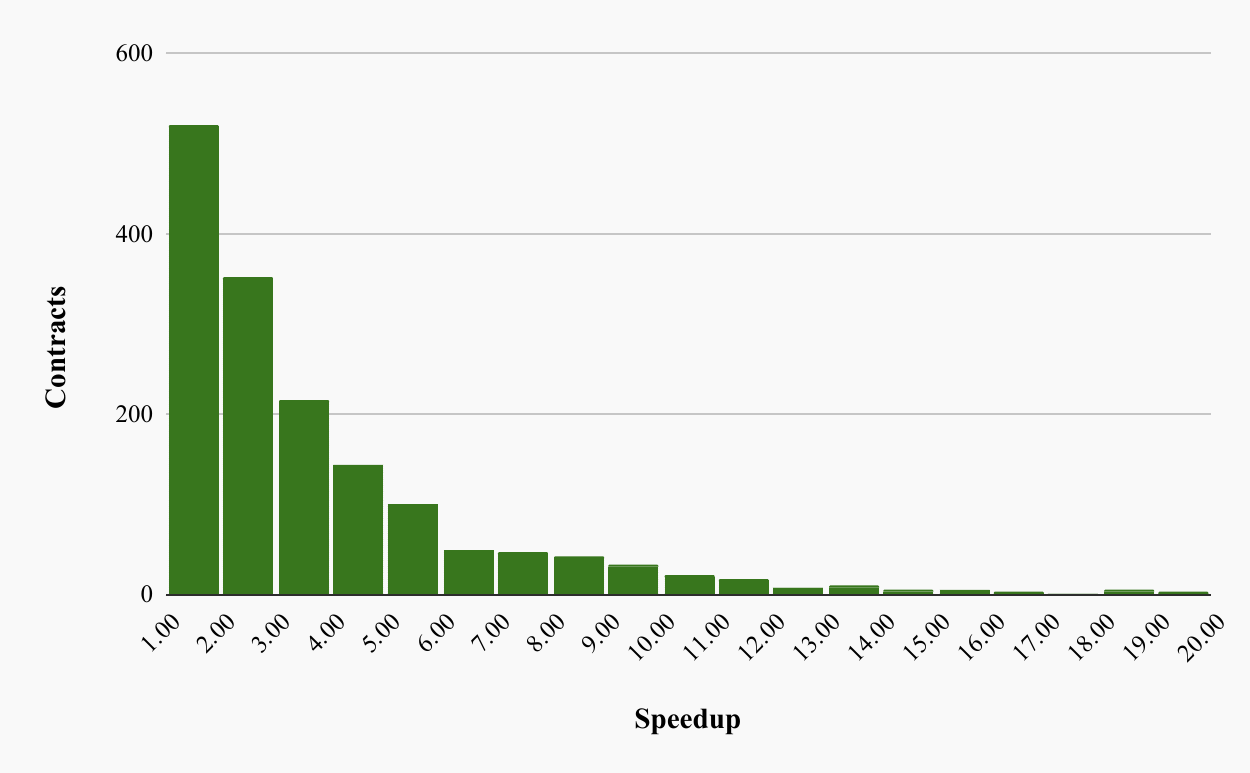}}
	\caption{\small Relative speedup due to value summary over a path-by-path function summary.
	}
	\label{fig:chart-speedup}
	\vspace{-7mm}
\end{figure}

\noindent
\textbf{Speedup due to \vsa:}
To characterize the performance gain from the \vsa, we have further designed this experiment where, instead of our value summary (\textbf{VS}), we provide a standard path-by-path function summary~\cite{solar,Godefroid07,AnandGT08} (\textbf{PS}) to the \refiner module.
From $\num{\clintStaticOnlyWarnings}$ contracts for which \toolname raised warnings (which are also the contracts sent to the \refiner), we randomly picked a subset of $\num{\clintSpeedupDataset}$ contracts 
\textbf{(i)} which belong to either medium, or large dataset, and 
\textbf{(ii)} \textbf{VS} configuration finished successfully without timing out---for this experiment.
We define \textit{speedup} factor $s = \frac{t_{ps}}{t_{vs}}$, where $t_m$ is the amount of time spent in the symbolic evaluation phase in mode $m$.
In \textbf{PS} mode, \toolname timed out for $\FPeval{\v}{round(\clintPathSummaryTimeout/\clintSpeedupDataset*100,2)}\v\%$ of the contracts owing to the increased cost of the \refine phase.
\Fig{fig:chart-speedup} presents a histogram of the speedup factor distribution of the remaining $\FPeval{\v}{round(\clintSpeedupDataset-\clintPathSummaryTimeout,0)}\num{\v}$ contracts for which the analyses terminated in both the modes.

\begin{mdframed}[style=graybox]
	Our novel value summary analysis is significantly faster than a classic summary-based analysis.
\end{mdframed}

\section{Case Studies}
\label{app:case_studies}
\subsection{Zero-day vulnerabilities}
\label{sec:case_study_zero_day}
\noindent
In this section, we present the unique vulnerabilities found by \toolname---not detected by any other tool.
We have redacted the code, and masked the program elements for the sake of anonymity and simplicity.
The fact that the origin of the \smart s can not be traced back in most of the cases makes it hard to report these bugs to the concerned developers.
Also, once a contract gets deployed, it is difficult to fix any bug due to the immutable nature of the blockchain.

\noindent
\textbf{Cross-function \reentrancy:}
\Fig{fig:case_study_dw_1} presents a simplified real-world contract---vulnerable to \textit{cross-function} \reentrancy attack due to Destructive Write (DW).
An attacker can set both \texttt{item\_1.creator} (\Line{11}), and \texttt{item\_1.game} (\Line{12}) to an arbitrary value by invoking \texttt{funcB()}.
In \texttt{funcA()}, an amount \texttt{amt} is transferred to \texttt{item\_1.creator} through \texttt{transferFrom}---an untrusted external contract call.
Therefore, when the external call is underway, the attacker can call \texttt{funcB()} to reset both \texttt{item\_1.creator}, and \texttt{item\_1.game}.
Hence, \texttt{item\_1.fee} gets transferred to a different address when \Line{6} gets executed.

\begin{figure}[h]
	\vspace{-5mm}
\begin{lstlisting}[
	caption=,
	stepnumber=1,
	firstnumber=1,
	basicstyle=\ttfamily\scriptsize,
%	linebackgroundcolor=
%	{\ifnum\value{lstnumber}=4\color{babypink}
%	\else\ifnum\value{lstnumber}=6\color{babypink}	
%	\else\fi\fi},
	language=Solidity]
function funcA(to, amt) public {
    ...
    IERC721 erc721 = IERC721(item_1.game)
    erc721.transferFrom(_, item1.creator, amt)
    ...
    item1.creator.transfer(item_1.fee)
}

function funcB(_creator, _game) {
    ...
    item_1.creator = _creator
    item1_1.game = _game
    ...
}
\end{lstlisting}
	\vspace{-7mm}
	\caption{Real-world cross-function \reentrancy
	}
	\label{fig:case_study_dw_1}
	\vspace{-3mm}
\end{figure}

\noindent
\textbf{Delegate-based \reentrancy:} \Fig{fig:case_study_delegate} presents a real-world contract, which is vulnerable to delegate-based \reentrancy attack.

\begin{figure}[h]
	\vspace{-4mm}
\begin{lstlisting}[
	caption=,
	stepnumber=1,
	firstnumber=1,
	basicstyle=\ttfamily\scriptsize,
%	linebackgroundcolor=
%	{\ifnum\value{lstnumber}=3\color{babypink}
%		\else\ifnum\value{lstnumber}=8\color{babypink}	
%		\else\fi\fi},
	language=Solidity]
function funcA(bytes _data) {
    __isTokenFallback = true;
    address(this).delegatecall(_data);
    __isTokenFallback = false;
}

function funcB(){
	assert(__isTokenFallback);
	// Write to application data
}

function funcC(address _to) {
    Receiver receiver = Receiver(_to)
    receiver.tokenFallback(...)
    ...
}

\end{lstlisting}
	\vspace{-6mm}
	\caption{Real-world delegatecall-based \reentrancy}
	\label{fig:case_study_delegate}
	\vspace{-3mm}
\end{figure}
The contract contains three functions---(a) \texttt{funcA} contains the \texttt{delegatecall},
(b) \texttt{funcB()} allows application data to be modified if the assertion is satisfied, and (c) \texttt{funcC} contains an untrusted external call.
A malicious payload can be injected in the \texttt{\_data} argument of \texttt{funcA}, which, in turn, invokes \texttt{funcC()} with a tainted destination \texttt{\_to}.
The \texttt{receiver} at \Line{14} is now attacker-controlled, which allows the attacker to reenter to \texttt{funcB} with \texttt{\_isTokenFallback} inconsistently set to \texttt{true}; thus rendering the assertion at \Line{8} useless.

\noindent
\textbf{CREAM Finance \reentrancy attack.}
By exploiting a \reentrancy vulnerability in the CREAM Finance, a decentralized lending protocol, the attacker stole $\num{462079976}$ AMP tokens, and $\num{2804.96}$ Ethers on August 31, 2021~\cite{cream}.
The attack involved two contracts: CREAM Finance contract \texttt{C}, and \texttt{AMP} token (ERC777) contract \texttt{A}.
The \texttt{borrow()} method of \texttt{C} calls the \texttt{transfer()} method of \texttt{A}, which, in turn, calls the \texttt{tokenReceived()} hook of the receiver contract \texttt{R}.
Such a hook is simply a function in \texttt{R} that is called when tokens are sent to it.
The vulnerability that the attacker leveraged is that there was a state (S) update in \texttt{C.borrow()} following the external call to \texttt{A.transfer()}.
Since, \texttt{A.transfer()} further calls \texttt{R.tokenReceived()} before even the original \texttt{C.borrow()} call returns, the attacker took this opportunity to reenter \texttt{C} before even the state update could take place.

Since the version of \slither that \toolname uses lacks support for all types of \solidity tuples, we could not run our tool as-is on the contract \texttt{C}.
To see whether our approach can still detect the above vulnerability by leveraging its inter-contract analysis, we redacted the contracts to eliminate syntactic complexity unrelated to the actual vulnerability.
When run on the simplified contract, \toolname successfully flagged it as vulnerable to the \reentrancy attack, as expected.

\noindent
\textbf{Transaction order dependency:} TOD may enable an attacker to earn profit by front-running a victim's transaction.
For example, during our manual analysis, we encountered a contract where the contract owner can set the price of an item on demand.
A user will pay a higher price for the item if the owner maliciously front-runs the user's transaction (purchase order), and sets the price to a higher value.
In another contract that enables buying and selling of tokens in exchange for \ether, the token price was inversely proportional with the current token supply.
Therefore, an attacker can front-run a buy transaction $T$, and buy $n$ tokens having a total price $p_l$.
After $T$ is executed, the token price will increase due to a drop in the token supply.
The attacker can then sell those $n$ tokens at a higher price, totaling price $p_h$, and making a profit of $(p_h - p_l)$.
We illustrate one more real-world example of a TOD attack in \Fig{fig:tod_impact} .
\begin{figure}[h]
	\vspace{-5mm}
\begin{lstlisting}[
	caption=,
	stepnumber=1,
	firstnumber=1,
	basicstyle=\ttfamily\scriptsize,
%	linebackgroundcolor=
%	{\ifnum\value{lstnumber}=7\color{babypink}
%		\else\fi},
	language=Solidity]
contract Bet {
 function recordBet(bool bet, uint _userAmount) {
  userBlnces[msg.sender]= _userAmount;
  totalBlnc[bet] = totalBlnc[bet] +_userAmount;    
 }
 function settleBet(bool bet) {
  uint reward = (userBlnces[msg.sender]*totalBlnc[!bet] 
		          / totalBlnc[bet];
  uint totalWth = reward + userBlnces[msg.sender];
  totalBlnc[!bet] = totalBlnc[!bet] - reward;
  msg.sender.transfer(totalWth);
 }
}
\end{lstlisting}
	\vspace{-0.22in}
	\caption{Real-world example of a TOD bug.
	}
	\label{fig:tod_impact}
	\vspace{-0.1in}
\end{figure}
\texttt{recordBet()} allows a user to place a bet, and then it adds (\Line{4}) the bet amount to the total balance of the contract.
In \texttt{settleBet()}, a user receives a fraction of the total bet amount as the reward amount.
Therefore, if two invocations of \texttt{settleBet()} having same \texttt{bet} value race against each other, the front-running one will earn higher reward as the value of \texttt{totalBlnc[!bet]}, which \texttt{reward} is calculated on, will also be higher in that case.

\subsection{Advantage of \vsa.}
\label{app:vsa_advantage}
\Fig{fig:case_study_value_summary} shows a real-world contract that demonstrates the benefit of the \vsa.
A \texttt{modifier} in \solidity{} is an additional piece of code which wraps the execution of a function.
Where the underscore (\_) is put inside the modifier decides when to execute the original function.
In this example, the public function \texttt{reapFarm} is guarded by the modifier \texttt{nonReentrant}, which sets the \texttt{reentrancy\_lock} (shortened as \texttt{L}) on entry, and resets it after exit.
Due to the \textit{hazardous access} (\Line{14} and \Line{18}) detected on \texttt{workDone}, \explorer flags this contract as potentially vulnerable.
However, the value summary analysis observes that the \texttt{require} clause at \Line{7} needs to be satisfied in order to be able to modify the lock variable \texttt{L}, which is encoded as: $\texttt{L} = \{\langle \mathtt{false},\mathtt{L=false} \rangle, \langle  \mathtt{true},\mathtt{L=false}  \rangle\}$.
In other words, there does not exist a program path that sets \texttt{L} to \texttt{false}, if the current value of \texttt{L} is \texttt{true}.
While making the external call at \Line{16}, the program state is $\valEnv=\{\mathtt{L}\mapsto \mathtt{true},...\}$, which means that \texttt{L} is \texttt{true} at that program point.
Taking both the value summary and the program state into account, the \refiner{} decides that the corresponding path leading to the \textit{potential} \reentrancy bug is infeasible.

\begin{figure}[t]
\begin{lstlisting}[
	caption=,
	stepnumber=1,
	firstnumber=1,
	basicstyle=\ttfamily\scriptsize,
%	linebackgroundcolor=
%	{\ifnum\value{lstnumber}=7\color{babypink}
%		\else\ifnum\value{lstnumber}=13\color{babypink}	
%		\else\fi\fi},
	language=Solidity]
interface Corn{
  function transfer(address to, uint256 value);
}
contract FreeTaxManFarmer {
  // Prevents re-entry to the decorated function
  modifier nonReentrant() {
    require(!reentrancy_lock);
    reentrancy_lock = true;
    _;
    reentrancy_lock = false;
  }

  function reapFarm(address tokn) nonReentrant {
    require(user[msg.sender][tokn].workDone > 0);
    // Untrusted external call
    Corn(tokn).transfer(msg.sender, ...);
    // State update
    user[msg.sender][tokn].workDone = 0;
  }
}
\end{lstlisting}
	\vspace{-5mm}
	\caption{The benefit of \vsa.
	}
	\label{fig:case_study_value_summary}
	%\vspace{-7mm}
\end{figure}

\subsection{False positives for \reentrancy and TOD}
\label{sec:case_study_reent_fp}
\noindent
\textbf{Reentrancy.} \Fig{fig:case_study_reentrancy_fp} features a real-world contract where \texttt{bTken} is set inside the constructor.
The static taint analysis that \toolname{} performs disregards the fact that \Line{5} is guarded by a \texttt{require} clause in the line before; thereby making the variable tainted.
Later at \Line{9} when the \texttt{balanceOf} method is invoked on \texttt{bTken}, \toolname raises a false alarm.

\begin{figure}[h]
	\vspace{-5.5mm}
\begin{lstlisting}[
	caption=,
	stepnumber=1,
	firstnumber=1,
	basicstyle=\ttfamily\scriptsize,
%	linebackgroundcolor=
%	{\ifnum\value{lstnumber}=4\color{babypink}
%	\else\fi},
	language=Solidity]
contract EnvientaPreToken {
  // Only owner can set bTken
  function enableBuyBackMode(address _bTken) {
    require( msg.sender == _creator );
    bTken = token(_bTken);
  }
  function transfer(address to, uint256 val) {
    // Trusted external call
    require(bTken.balanceOf(address(this))>=val);
    balances[msg.sender] -= val;
  }
}
\end{lstlisting}
	\vspace{-6mm}
	\caption{False positive of \toolname (Reentrancy).
		}
	\label{fig:case_study_reentrancy_fp}
	\vspace{-4mm}
\end{figure}

\noindent
\textbf{TOD.} \Fig{fig:case_study_tod_fp} presents a real-world donation collection contract, where the contract transfers the collected donations to its recipient of choice.
Both \toolname and \securify raised TOD warning as the transferred amount, \ie, \texttt{donations} at \Line{7}, can be modified by function \texttt{pay()} at \Line{3}.
Though the amount of \ether withdrawn (\texttt{donations}) is different depending on which of \texttt{withdrawDonations()} and \texttt{pay()} get scheduled first---this does not do any harm as far as the functionality is concerned.
In fact, if \texttt{pay()} front-runs \texttt{withdrawDonations()}, the recipient is rewarded with a greater amount of donation.
Therefore, this specific scenario does not correspond to a TOD attack. 

\begin{figure}[h]
	\vspace{-5.5mm}
\begin{lstlisting}[
	caption=,
	stepnumber=1,
	basicstyle=\ttfamily\scriptsize,
	firstnumber=1,
	language=Solidity]
contract Depay{
    function pay(..., uint donation) {
        donations += donation;
    }
    function withdrawDonations(address recipient) {
    	require(msg.sender == developer)
        recipient.transfer(donations);
    }
}
\end{lstlisting}
	\vspace{-0.22in}
	\caption{False positive of TOD.
		}
	\label{fig:case_study_tod_fp}
	\vspace{-5mm}
\end{figure}

\section{Extended Related Work}
\label{app:extended_related_work}
\noindent
\textbf{Hybrid analysis.}
Composition of static analysis and symbolic execution has been applied to find bugs in programs other than \smart s.
For example, \sys~\cite{Fraser20} uses static analysis to find potential buggy paths in large codebases, followed by an under-constrained symbolic execution to verify the feasibility of those paths.
\woodpecker~\cite{Heming13} uses rule-directed symbolic execution to explore only relevant paths in a program.
To find double fetch bugs in OS kernels, \deadline~\cite{Meng18} employs static analysis to prune paths, and later performs symbolic execution only on those paths containing multiple reads.
Several other tools~\cite{Domagoj11,Josselin16, Gerasimov18,Shengjian15,Shengjian16} employ similar hybrid techniques for testing, verification, and bug finding.
Such hybrid analyses have been proved effective to either prune uninteresting paths, or selectively explore interesting parts of the program.
In \toolname, we use static analysis to filter out interesting contracts, find potentially vulnerable paths, and compute value-summary to be used in conjunction with the symbolic execution---to achieve both scalability, and precision.

\section{Extended Discussion}
\label{app:discussion}

\noindent
\textbf{Imprecise analysis components.}
\toolname{} performs inter-contract analysis (\Appen{app:inter_contract_analysis}) when the source code of the called contract is present in our database, and more importantly, the external call destination $d$ is statically known.
If either of the conditions does not hold, \toolname treats such an external call as \textit{untrusted}, thereby losing precision.
The question of external call destination $d$ resolution comes only when \toolname is used on contracts that have been deployed already.
For cases where $d$ is set at run-time, our prototype relies on only contract creation transactions.
If $d$ is set through a public setter method, our current prototype cannot detect those cases, though it would not be hard to extend the implementation to support this case as well.
Moreover, \toolname incurs false positives due to the imprecise taint analysis engine from \slither.
Therefore, using an improved taint analysis will benefit \toolname's precision.

\noindent
\textbf{Bytecode-based analysis.}
\toolname relies on control-flow recovery, taint analysis, and symbolic evaluation as its fundamental building blocks.
Recovering source-level rich data structures, \eg, array, strings, mappings, \etc, is not a requirement for our analysis.
Even for EVM bytecode, recovering the entry points of public methods is relatively easier due to the ``jump-table'' like structure that the \solidity compiler inserts at the beginning of the compiled bytecode.
Typically, it is expected for a decompiler platform to provide the building blocks in the form of an API, which then could be used to port \toolname for bytecode analysis.
That said, the performance and precision of our analysis are limited by the efficacy of the underlying decompiler.
Thanks to the recent research~\cite{rattle-repo,gigahorse,panoramix-repo,Lagouvardos20} on EVM decompilers and static analysis, significant progress has been made in this front.

\noindent
\textbf{Other bugs induced by \haz.}
If a contract contains hazardous access, but no reentrancy/TOD vulnerability, that can still lead to a class of bugs called Event Ordering (EO) bugs~\cite{ethracer}, due to the asynchronous callbacks initiated from an off-chain service like \texttt{Oraclize}.
We consider such bugs as out of scope for this work.

\section{Technical details}
\label{app:technical_details}

\subsection{Inter-contract analysis}
\label{app:inter_contract_analysis}
To model inter-contract interaction as precisely as possible, we perform a backward data-flow analysis starting from the destination $d$ of an external call (\eg, $\mathsf{call}$, $\mathsf{delegatecall}$, \etc), which leads to the following three possibilities:
\textbf{(a)} $d$ is visible from source,
\textbf{(b)} $d$ is set by the \textit{owner} at run-time, \eg, in the constructor during contract creation.
In this case, we further infer $d$ by analyzing existing transactions, \eg, by looking into the arguments of the contract-creating transaction, and
\textbf{(c)} $d$ is attacker-controlled.
While crawling, we build a database from the contract address to its respective source.
Hence, for cases (a) and (b) where $d$ is statically known, we incorporate the target contract in our analysis if its source is present in our database.
If either the source is not present, or $d$ is tainted (case (c)), we treat such calls as \textit{untrusted}, requiring no further analysis.\looseness=-1

\subsection{Detecting \textit{owner-only} statements}
\label{app:owner_only_statements}
In the context of \smart, the \textit{owner} refers to one or more addresses that play certain administrative roles, \eg, contract creation, destruction, \etc{}
Typically, critical functionalities of the contract can only be exercised by the owner.
We call the statements that implement such functionalities as \textit{owner-only} statements.
Determining the precise set of owner-only statements in a contract can be challenging as it requires reasoning about complex path conditions.
\toolname, instead, computes a over-approximate set of owner-only statements during the computation of base ICFG facts.
This enables \toolname, during the \explore phase, not to consider certain \haz pairs that can not be exercised by an attacker.
To start with, \toolname initializes the analysis by collecting the set of storage variables (owner-only variables) $\mathcal{O}$ defined during the contract creation.
Then, the algorithm computes the transitive closure of all the storage variables which have \textit{write} operations that are control-flow dependent on $\mathcal{O}$.
Finally, to compute the set of owner-only statements, \toolname collects the statements which have their execution dependent on $\mathcal{O}$.

\end{document}